\documentclass{emulateapj}

\usepackage{graphicx} 
\usepackage{ifpdf}

\pdfoutput =  0

\usepackage{amsmath}
\newcommand{\numax}{\mbox{$\nu_{\rm max}$}}
\newcommand{\Dnu}{\mbox{$\Delta \nu$}}

\newcommand{\muHz}{\mbox{$\mu$Hz}}
\newcommand{\kep}{\mbox{\textit{Kepler}}}

\newcommand{\msun}{\mbox{$M_{\sun}$}}
\newcommand{\rsun}{\mbox{$R_{\sun}$}}

\slugcomment{accepted for publication in ApJ}

\shorttitle{Fundamental Properties of Stars using Asteroseismology and 
Interferometry}
\shortauthors{D. Huber et al.}

\def\apjl{ApJ}%
\def\apjs{ApJS}%
\def\ao{Appl.~Opt.}%
\def\aap{A\&A}%
\begin{document}

\title{Fundamental Properties of Stars using Asteroseismology from \textit{KEPLER} \& CoRoT and 
Interferometry from the CHARA Array}

\author{
D.~Huber\altaffilmark{1,2,3}, 
M.~J.~Ireland\altaffilmark{1,4,5}, 
T.~R.~Bedding\altaffilmark{1}, 
I.~M.~Brand\~ao\altaffilmark{6}, 
L.~Piau\altaffilmark{7}, 
V.~Maestro\altaffilmark{1}, 
T.~R.~White\altaffilmark{1}, 
H.~Bruntt\altaffilmark{8}, 
L.~Casagrande\altaffilmark{9}, 
J.~Molenda-\.Zakowicz\altaffilmark{10}, 
V.~Silva Aguirre\altaffilmark{8}, 
S.~G.~Sousa\altaffilmark{6}, 
T.~Barclay\altaffilmark{11}, 
C.~J.~Burke\altaffilmark{12}, 
W.~J.~Chaplin\altaffilmark{13,8}, 
J.~Christensen-Dalsgaard\altaffilmark{8}, 
M.~S.~Cunha\altaffilmark{6}, 
J.~De~Ridder\altaffilmark{14}, 
C.~D.~Farrington\altaffilmark{15}, 
A.~Frasca\altaffilmark{16}, 
R.~A.~Garc\'\i a\altaffilmark{17}, 
R.~L.~Gilliland\altaffilmark{18}, 
P.~J.~Goldfinger\altaffilmark{15}, 
S.~Hekker\altaffilmark{19}, 
S.~D.~Kawaler\altaffilmark{20}, 
H.~Kjeldsen\altaffilmark{8}, 
H.~A.~McAlister\altaffilmark{15}, 
T.~S.~Metcalfe\altaffilmark{21}, 
A.~Miglio\altaffilmark{13}, 
M.~J.~P.~F. G. Monteiro\altaffilmark{6}, 
M.~H.~Pinsonneault\altaffilmark{22}, 
G.~H.~Schaefer\altaffilmark{15}, 
D.~Stello\altaffilmark{1}, 
M.~C.~Stumpe\altaffilmark{12}, 
J.~Sturmann\altaffilmark{15}, 
L.~Sturmann\altaffilmark{15}, 
T.~A.~ten Brummelaar\altaffilmark{15}, 
M.~J.~Thompson\altaffilmark{23}, 
N.~Turner\altaffilmark{15}, and 
K.~Uytterhoeven\altaffilmark{24,25} 
}
\altaffiltext{1}{Sydney Institute for Astronomy (SIfA), School of Physics, University of Sydney, NSW 2006, Australia}
\altaffiltext{2}{NASA Ames Research Center, MS 244-30, Moffett Field, CA 94035, USA}
\altaffiltext{3}{NASA Postdoctoral Program Fellow; \mbox{daniel.huber@nasa.gov}}
\altaffiltext{4}{Department of Physics and Astronomy, Macquarie University, NSW 2109, Australia}
\altaffiltext{5}{Australian Astronomical Observatory, PO Box 296, Epping, NSW 1710, Australia}
\altaffiltext{6}{Centro de Astrof\'{\i}sica and Faculdade de Ci\^encias, Universidade do Porto, Rua das Estrelas, 4150-762 Porto, Portugal}
\altaffiltext{7}{Department of Physics and Astronomy, Michigan State University, East Lansing, MI 48823-2320, USA}
\altaffiltext{8}{Stellar Astrophysics Centre, Department of Physics and Astronomy, Aarhus University, Ny Munkegade 120, DK-8000 Aarhus C, Denmark}
\altaffiltext{9}{Research School of Astronomy \& Astrophysics, Mount Stromlo Observatory, The Australian National University, ACT 2611, Australia}
\altaffiltext{10}{Astronomical Institute of the University of Wroc{\l}aw, ul. Kopernika 11, 51-622 Wroc{\l}aw, Poland}
\altaffiltext{11}{Bay Area Environmental Research Institute/NASA Ames Research Center, Moffett Field, CA 94035}
\altaffiltext{12}{SETI Institute/NASA Ames Research Center, MS 244-30, Moffett Field, CA 94035, USA}
\altaffiltext{13}{School of Physics and Astronomy, University of Birmingham, Birmingham B15 2TT, UK}
\altaffiltext{14}{Instituut voor Sterrenkunde, K.U.Leuven, Belgium}
\altaffiltext{15}{Center for High Angular Resolution Astronomy, Georgia State University, PO Box 3969, Atlanta, GA 30302, USA}
\altaffiltext{16}{INAF Osservatorio Astrofisico di Catania, Italy}
\altaffiltext{17}{Laboratoire AIM, CEA/DSM-CNRS, Universit\'e Paris 7 Diderot, IRFU/SAp, Centre de Saclay, 91191, Gif-sur-Yvette, France}
\altaffiltext{18}{Space Telescope Science Institute, 3700 San Martin Drive, Baltimore, Maryland 21218, USA}
\altaffiltext{19}{Astronomical Institute 'Anton Pannekoek', University of Amsterdam, Science Park 904, 1098 XH Amsterdam, The Netherlands}
\altaffiltext{20}{Department of Physics and Astronomy, Iowa State University, Ames, IA 50011 USA}
\altaffiltext{21}{Space Science Institute, Boulder, CO 80301, USA}
\altaffiltext{22}{Department of Astronomy, Ohio State University, OH 43210, USA}
\altaffiltext{23}{High Altitude Observatory, NCAR, P.O. Box 3000, Boulder, CO 80307, USA}
\altaffiltext{24}{Instituto de Astrofisica de Canarias, 38205 La Laguna, Tenerife, Spain}
\altaffiltext{25}{Dept. Astrof\'{\i}sica, Universidad de La Laguna (ULL), Tenerife, Spain}

\begin{abstract}
We present results of a long-baseline interferometry campaign using the PAVO beam combiner at 
the CHARA Array to measure 
the angular sizes of five main-sequence stars, one subgiant and four red giant stars for 
which solar-like oscillations have been detected by either \kep\ or CoRoT. By combining 
interferometric angular diameters, Hipparcos parallaxes, asteroseismic densities, bolometric 
fluxes and high-resolution spectroscopy we 
derive a full set of near model-independent fundamental properties for the sample.
We first use these properties to test asteroseismic 
scaling relations for the frequency of maximum power (\numax) and the large frequency separation 
(\Dnu). We find excellent agreement within the observational uncertainties, and empirically show 
that simple estimates of asteroseismic radii for main-sequence stars are accurate to $\lesssim 4\%$. 
We furthermore find good agreement 
of our measured effective temperatures with spectroscopic and photometric estimates 
with mean deviations for stars between $T_{\rm eff} = 4600-6200$\,K of $-22\pm32$\,K 
(with a scatter of 97\,K) and 
$-58\pm31$\,K (with a scatter of 93\,K), respectively.
Finally we present a first comparison 
with evolutionary models, and find differences between observed and 
theoretical properties for the metal-rich main-sequence star HD\,173701.
We conclude that the constraints presented in this study
will have strong potential for testing stellar model physics, in particular when combined with 
detailed modelling of individual oscillation frequencies.
\end{abstract}


\section{Introduction}
The knowledge of fundamental properties such as temperature, radius and 
mass of stars in different evolutionary phases plays a key role in many applications of 
modern astrophysics. Examples include the improvement of model physics of stellar 
structure and evolution such as convection 
\citep[see, e.g.][]{demarque86,monteiro96,deheuvels11,trampedach11,piau11}, 
the calibration of empirical relations 
such as the color-temperature scale for cool stars 
\citep[see, e.g.,][]{flower96,ramirez05,casagrande10}, and 
the characterization of physical properties and habitable zones of exoplanets 
\citep[see, e.g.,][]{baines08,vanbelle09,vonbraun11b,vonbraun11}.

Many methods to determine properties of single field stars are indirect, and therefore 
of limited use for improving stellar models. 
Asteroseismology, the study of stellar oscillations, is
a powerful method to determine 
properties of solar-type stars such as the mean stellar density with little model dependence
\citep[see, e.g.,][]{brown94, CD04, aerts10}. 
Additionally, long-baseline interferometry 
can be used to measure the angular sizes of stars which, in combination with a  
parallax, yields a linear radius and, when combined with an estimate for the bolometric flux, 
provides a direct measurement of a star's effective temperature 
\citep[see, e.g.,][]{code,boyajian09,baines09,boyajian12,boyajian12b,creevey12b}. 
Therefore, the combination of both 
methods in principle allows a determination of radii, masses and temperatures of stars
with little model dependence.

While the potential of combining 
asteroseismology and interferometry has been long recognized \citep[see, e.g,][]{cunha07}, 
observational constraints have so far restricted 
an application for cool stars to relatively few bright objects 
\citep{north07,bruntt10,bazot11}. Recent technological 
advances, however, have changed this 
picture. The launches of the space telescopes CoRoT 
\citep[\textit{Convection, Rotation and planetary Transits},][]{baglin06,baglin06b} 
and \kep\ \citep{borucki10,koch10b} has increased the number of stars 
with detected solar-like oscillations to several thousands, providing a large 
sample spanning from the main-sequence to He-core burning red giant stars 
\citep{deridder09, hekker09, gilliland10,chaplin11a}. 
At the same time, the development of highly sensitive instruments such as the PAVO 
beam combiner \citep{ireland08} at the CHARA Array \citep{brummelaar04} have pushed the 
sensitivity limits of long-baseline 
interferometry, bringing into reach the brightest objects for which high-quality space-based 
asteroseismic data are available. Using these recent advances, we present a systematic 
combined asteroseismic and 
interferometric study of low-mass stars spanning from the main-sequence to the red clump.

\section{Target Sample}
Our target sample was selected to optimize the combination of asteroseismology and interferometry  
given the observational constraints, while also covering a large parameter space in 
stellar evolution. The majority of our stars were taken from the sample analyzed by the 
Kepler Asteroseismic Science Consortium (KASC). We selected four unevolved stars, 
which are among the brightest oscillating solar-type stars observed by \kep. 
Note that our interferometric results for
$\theta$\,Cyg \citep{guzik11} and 16\,Cyg A\&B \citep{metcalfe12} 
will be presented elsewhere. For the \kep\ giant sample, four of the brightest 
red giants with the best Hipparcos parallaxes were selected. Finally, the main-sequence stars 
HD\,175726 and HD\,181420 in our sample are located in the CoRoT field towards the 
galactic center, and were among the first CoRoT main-sequence stars with detected oscillations 
\citep{barban09,mosser09c}. Note that our PAVO campaign is also targeting solar-type oscillators 
in the CoRoT field in the galactic anti-center, such as the F-star 
HD\,49933 \citep{appourchaux08,benomar09,kallinger10b}, which has already 
been subject to interferometric follow-up \citep{bigot11}. However, due to poor weather 
conditions during the winter seasons on Mt.\,Wilson, not enough data has yet been collected for these 
targets.

\begin{figure}
\begin{center}
\resizebox{\hsize}{!}{\includegraphics{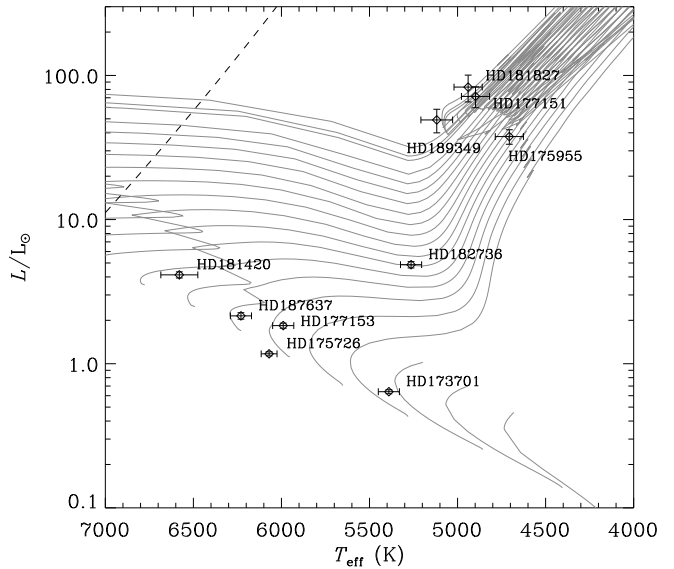}}
\caption{H-R diagram with the position of the target 
stars calculated using spectroscopy, photometry and 
Hipparcos parallaxes. Solar metallicity BaSTI evolutionary tracks from 0.6-2.6\,$M_{\sun}$ in 
steps of 0.1\,$M_{\sun}$ are shown as grey lines. The dashed line marks the approximate location of 
the cool edge of the instability strip.}
\label{fig:hrd2}
\end{center}
\end{figure}

In the remainder of this section we summarize the basic parameters of our target stars derived
using classical methods and measurements available in the literature.
Table \ref{tab:litparas} lists the complete target sample of our study, with spectral types 
taken from the HD catalog. Nine of the ten stars in our sample have 
atmospheric parameters derived from modeling several hundred lines in high-resolution 
spectra using the VWA package \citep{bruntt10}, as presented by \citet{bruntt09}, \citet{bruntt12} 
and \citet{thygesen12}. These are also listed in Table \ref{tab:litparas}.
For HD\,189349, 
we have analyzed a spectrum obtained with the NARVAL spectrograph at the Pic du 
Midi Observatory using three different methods: VWA \citep{bruntt10}, ROTFIT 
\citep{frasca03,frasca06} and the method described by \citet{santos04}, \citet{sousa06} and 
\citet{sousa08}. In two of the three methods the surface gravity was fixed to the value calculated 
from asteroseismic scaling relations (see next section). The resulting spectroscopic parameters  
for each method are listed in Table \ref{tab:litparas2}, and we have adopted a weighted mean of all three methods, 
given in Table \ref{tab:litparas}, for the remainder of this paper.
Note that for HD\,173701 spectroscopic 
parameters have also been published by \citet{valenti05}, \citet{mishenina04} and \citet{kovtyukh03}, 
which are also listed in Table \ref{tab:litparas2} for comparison. The published values are in good 
agreement with the values adopted here.

\begin{table*}
\begin{center}
\caption{Fundamental properties of target stars using available literature 
information. Stars are separated into main-sequence and subgiant stars (top) and red giants (bottom).}
\vspace{0.1cm}
\begin{small}
\begin{tabular}{l l c c c c c c c c}        
\hline         
HD  & KIC &	Sp.T. & $V$ & $B-V$ 	 & \multicolumn{3}{c}{Spectroscopy}	 &  \multicolumn{2}{c}{Hipparcos}	\\ 
    &	  &    	      &     &       & $T_{\rm eff}$ & $\log g$              & [Fe/H] &  $\pi$ (mas)   & $L/L_{\sun}$ \\
\hline
173701$^{1}$&	8006161	&  K0V   &  $7.514$ & $0.878$ & 5390(60)  & 4.49(3) & $+0.34$(6) & 37.47(49) & 0.64(2)	 \\
175726$^{2}$&	--	    &  G5V   &  $6.711$ & $0.571$ & 6070(45)  & 4.53(4) & $-0.07$(3) & 37.73(51) & 1.17(4)	 \\
177153$^{1}$&   6106415 &  G0V   &  $7.205$ & $0.558$ & 5990(60)  & 4.31(3) & $-0.09$(6) & 24.11(44) & 1.84(7)	 \\
181420$^{2}$&	--	    &  F2V   &  $6.561$ & $0.434$ & 6580(105) & 4.26(8) & $+0.00$(6) & 21.05(48) & 4.1(2)	  \\
182736$^{1}$&	8751420	&  G0IV  &  $7.022$ & $0.800$ & 5264(60)  & 3.70(3) & $-0.15$(6) & 17.35(41) & 4.9(2)	  \\
187637$^{1}$&	6225718	&  F5V 	 &  $7.520$ & $0.500$ & 6230(60)  & 4.32(3) & $-0.17$(6) & 19.03(46) & 2.1(1)	  \\
\hline
175955$^{3}$&	10323222&  K0III &  $7.014$ & $1.171$ & 4706(80) & 2.60(1)  & $+0.06$(15) & 7.62(38) &  38(4)		  \\
177151$^{3}$&	10716853&  K0III &  $7.040$ & $0.994$ & 4898(80) & 2.62(1)  & $-0.10$(15) & 4.92(38) &  72(12)		  \\
181827$^{3}$&	8813946	&  K0III &  $7.188$ & $1.012$ & 4940(80) & 2.81(1)  & $+0.14$(15) & 4.23(43) &  83(18)		  \\
189349      &	5737655	&  G5III &  $7.305$ & $0.878$ & 5118(90) & 2.4(1)   & $-0.56$(16)  & 5.32(47) &  49(9)	     \\ 							
\hline
\end{tabular} 
\label{tab:litparas} 
\flushleft $B$ and $V$ magnitudes are Tycho photometry \citep{perryman97} converted 
into the Johnson system using the calibration by \citet{bessell00}. 
Spectroscopic parameters were adopted from $^{1}$\citet{bruntt12}, $^{2}$\citet{bruntt09} 
and $^{3}$\citet{thygesen12}. Spectroscopic parameters for HD\,189349 are the 
weighted average of three results presented in this work (see text and 
Table \ref{tab:litparas2}). Brackets indicate the uncertainties on a parameter (note that 
this notation has been adopted throughout the paper).
\end{small}
\end{center}
\end{table*}

All stars in our sample have measured 
Hipparcos parallaxes \citep{leeuwen07}, with uncertainties ranging from $\sim$ 1 to 10\,\%.
All unevolved stars in our sample are at distances 
$<60$\,pc and hence reddening is expected to be negligible \citep[see][]{molenda09,bruntt12}. 
Hence, we assumed zero 
reddening for all unevolved stars with an uncertainty of 0.005\,mag. 
For the giants, we have estimated reddening by comparing observed colors 
to synthetic photometry of models matching the spectroscopic parameters in Table \ref{tab:litparas}, 
as described in more detail in Section \ref{sec:bolflux}. To estimate an 
uncertainty, we have compared these values to $E(B-V)$ values listed in the 
Kepler Input Catalog \citep[KIC, ][]{brown11} for 
nearby stars and to estimates from the 3-D extinction model by \citet{drimmel03}. 
The mean scatter between these methods for all stars is 0.02\,mag, which we adopt as 
our uncertainty in $E(B-V)$ for the giants in our sample.
Finally, we used the spectroscopically determined 
effective temperatures and metallicities to estimate a bolometric 
correction 
for each star using the calibrations by \citet{flower96} and \citet{alonso99} with appropriate 
zero-points as discussed in \citet{torres10}, yielding the stellar luminosity given in the 
last column of Table \ref{tab:litparas}.
Figure 1 shows an H-R diagram of our target stars, according to the properties listed in Table 
\ref{tab:litparas}, together with solar-metallicity BaSTI 
evolutionary tracks \citep{basti}.

\begin{table}
\begin{small}
\begin{center}
\caption{Atmospheric parameters for stars from different methods.}
\vspace{0.1cm}
\begin{tabular}{l l c c c l}        
\hline         
HD & KIC & $T_{\rm eff}$ & $\log g$ & [Fe/H] & Ref \\
\hline         
173701 & 8006161 & 5390(60) & 4.49(3) 	& 0.34(6) & 1$^{*}$   \\
&	& 5399(44) & 4.53(6) 		& 0.32(3) & 2	   \\
&  	& 5423(20) & 4.4(2)		& 0.2(1)  & 3	   \\
&  	& 5423(10) & -- & -- & 4     \\
\hline
189349 & 5737655  & 5070(100) & 2.4(1)	  & -0.7(1)  & 5a$^{*}$       \\
&	& 5145(63)  & 2.4(1)	  & -0.54(5)  & 5b$^{*}$       \\
&	& 5163(71)  & 2.9(2)		& -0.44(11) & 5c \\
\hline 	    
\end{tabular} 
\label{tab:litparas2} 
\end{center}
\flushleft $^{*} \log g$ fixed to asteroseismic value. (1) \citet{bruntt12}, 
(2) \citet{valenti05}, (3) \citet{mishenina04}, (4) \citet{kovtyukh03}, 
(5) this paper: a - VWA \citep{bruntt10}, 
b - \citet{santos04}, \citet{sousa06} and 
\citet{sousa08}, c - ROTFIT \citep{frasca03,frasca06}.
\end{small}
\end{table}

\section{Observations}

\subsection{Asteroseismology}
The asteroseismic results presented in this paper are based on observations obtained by the \kep\ and 
CoRoT space telescopes. Both satellites deliver near-uninterrupted, high S/N time series which 
are ideally suited for asteroseismic studies. In this paper, we focus on 
two global parameters: the frequency of maximum power (\numax) and the 
large frequency separation (\Dnu). These are frequently used 
to determine fundamental properties of main-sequence and red giant stars 
\citep[see, e.g.,][]{stello09,kallinger10,kallinger10c,chaplin11a,hekker11b,hekker11c,silvaaguirre11,huber11b}. 
For a general introduction to solar-like oscillations, we refer the reader to the review 
by \citet{bedding11b}.

\begin{figure*}
\begin{center}
\resizebox{\hsize}{!}{\includegraphics{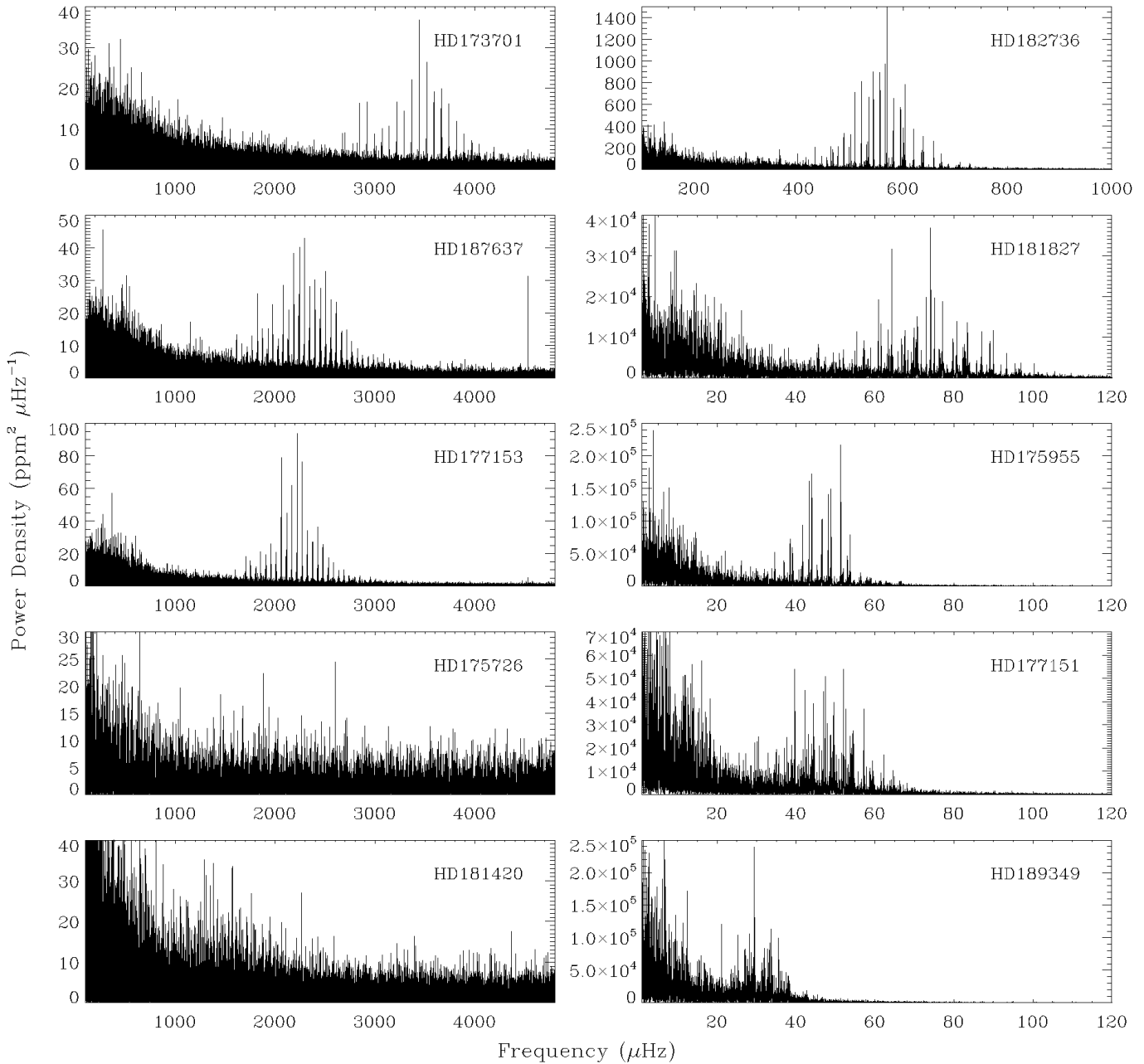}}
\caption{Power density spectra for all stars in our sample, sorted by the frequency 
of maximum power (\numax). Note the change in x-axis scale for main-sequence (left column), 
subgiant (top right column) and red-giant stars (four bottom right panels). Note that the 
high peak at $\sim$\,4500\muHz\ for HD\,187637 is a known artefact of \kep\ short-cadence data 
\citep{gilliland10b}.}
\label{fig:ps_chara}
\end{center}
\end{figure*}

Figure \ref{fig:ps_chara} presents the power spectrum for each star, sorted by 
the frequency of maximum power (\numax). In most cases, a clear power excess 
due to solar-like oscillations is visible. A 
summary of the datasets used in our analysis, as well as the derived asteroseismic parameters, 
is given in Table \ref{tab:seism}.
The analysis of \kep\ stars is based on either short-cadence \citep{gilliland10b} 
or long-cadence 
\citep{jenkins10} data up to Q10, which were corrected for instrumental trends as described in \citet{garcia11}.
Global asteroseismic parameters were extracted using the automated analysis pipeline by 
\citet{huber09}, which has been shown to agree well with other methods 
\citep{hekker11,verner11}. 
Due to the length and very high S/N of the \kep\ data, the modes are 
resolved and uncertainties on \numax\ and (particularly) \Dnu\ are
dominated by the adopted method (e.g., the range over which \Dnu\ is determined) 
rather than measurement errors. To account for this, we 
added in quadrature to the formal uncertainties 
an uncertainty based on the scatter of different methods 
used by \citet{silva12} for short-cadence data and by \citet{huber11b} for long-cadence data. 
The analysis by \citet{huber11b} 
was based on data spanning from Q0-6, which in most cases was sufficient to 
resolve the modes and reliably estimate \numax\ and \Dnu\ \citep{hekker12}.
In general, the uncertainties on the asteroseismic parameters for most \kep\ stars are 
negligible compared to the uncertainties on other observables.
A notable exception is HD\,189349, with a relatively large uncertainty of $\sim 4$\% in the 
large frequency separation. Inspection of the power spectrum shows that 
the modes for this star are very broad, making a determination of $\Dnu$ difficult. 
We speculate that the unusually broad modes 
may be related to the low metallicity of this object, 
but a more in-depth analysis is beyond the scope of this paper.

\begin{table*}
\begin{center}
\caption{Asteroseismic observations and measured parameters.}
\vspace{0.1cm}
\begin{small}
\begin{tabular}{l l l c c c c}        
\hline         
HD  & KIC  &Data	&	T (d)	& Duty cycle (\%)& \numax (\muHz)	& \Dnu (\muHz) 	\\ 
\hline
173701& 8006161  & Kepler SC Q5-10     &  557  & 92  &  3619(98)	&  149.3(4)	     \\
175726$^{1}$& --	 & CoRoT SRc01     &   28  & 90  &  1915(200)	&   97.2(5)     \\
177153& 6106415  & Kepler SC Q6-8,10   &  461  & 70  &  2233(60)	&  104.3(3)	   \\
181420& --	 & CoRoT LRc01             &  156  & 90  &  1574(31)	&   75.1(3)	   \\
182736& 8751420  & Kepler SC Q5,7-10   &  557  & 75  &   568(15)	&   34.6(1)	    \\
187637& 6225718  & Kepler SC Q6-10     &  461  & 91  &  2352(66)	&  105.8(3)	    \\
\hline
175955& 10323222 & Kepler LC Q0-10     &  880  & 91  &  46.7(1.1)	&  4.86(3)	       \\
177151& 10716853 & Kepler LC Q1-7,9-10 &  869  & 77  &  48.8(1.1)	&  4.98(7)	       \\
181827& 8813946  & Kepler LC Q1-10     &  869  & 91  &  73.1(1.2)	&  6.45(7)	\\
189349& 5737655  & Kepler LC Q1-10     &  869  & 91  &  29.9(1.1)	&  4.22(16)		 \\		   
\hline
\end{tabular} 
\label{tab:seism}
\flushleft $^{1}$Detection adopted from \citet{mosser09c}.
\end{small}
\end{center}
\end{table*}

For the two CoRoT stars in our sample, we have re-analyzed publicly available data 
using the method described in \citet{huber09}. Our results for HD181420 are in 
good agreement with the 
values published by \citet{barban09}. For HD175726, our 
analysis did not yield significant evidence for regularly spaced peaks, and yielded only 
marginal evidence for a power excess at $1900\pm200\muHz$. 
\citet{mosser09c} have argued that this power excess is compatible with solar-like oscillations and 
showed evidence for a 
large variation of \Dnu\ with frequency, which could be responsible for the null-detection 
in our analysis.
We have adopted the 
published value for \Dnu\ by \citet{mosser09c} and a value for \numax\ corresponding to the 
maximum of the power excess in the spectrum, with a conservative uncertainty of 10\%.

\subsection{Interferometry}

Interferometric observations were made with the Precision Astronomical Visible Observations 
(PAVO) beam combiner \citep{ireland08} at the Center for High Angular Resolution Astronomy (CHARA) 
on Mt. Wilson observatory, California 
\citep{brummelaar04}. Operating at a central wavelength of $\lambda=0.7\,\mu$m with baselines 
up to 330\,m, PAVO at CHARA is one of the 
highest angular-resolution instruments world-wide.

A complete description of the instrument was given by \citet{ireland08}, and we summarize the basic 
aspects here. The light from up to three telescopes passes through vacuum tubes and into 
a series of optics to 
compensate the path difference. The beams are then collimated and passed through a non-redundant mask 
which acts as a bandpass filter, and spatially modulated interference fringes are formed behind the mask. 
The interference pattern is then passed through a lenslet array and a 
prism, producing fringes in 16 segments on the CCD detector, each being spectrally 
dispersed in several independent wavelength channels. Major 
advantages of the PAVO design are high sensitivity (with a
limiting magnitude of $R\sim 8$\,mag in typical seeing conditions),
increased information through spectral dispersion, and high spatial resolution through 
operating at visible wavelengths. 
First PAVO science results have been presented by \citet{bazot11},
\citet{derekas11} and \citet{huber12}. For our analysis, we have used PAVO 
observations in two-telescope mode, with baselines ranging from $\sim 110-330$\,m. 

\begin{table}
\begin{small}
\begin{center}
\caption{Calibrators used for interferometric observations.}
\vspace{0.1cm}
\begin{tabular}{l c c c c c}        
\hline         
HD  	& Sp.T.	& $V-K$ &  $E(B-V)$   & $\theta_{V-K}$ & ID \\  
\hline
171654  &   A0V   &	 -0.067  &    0.036  &   0.141 & c    \\
174177  &   A0V   &	  0.249  &    0.020  &   0.191 & gh   \\
176131  &   A2V   &	  0.345  &    0.012  &   0.155 & ac   \\
176626  &   A2V   &	  0.084  &    0.026  &   0.146 & ac   \\
177959  &   A3V   &	  0.451  &    0.029  &   0.152 & b    \\
178190  &   A2V   &	  0.381  &    0.027  &   0.157 & bd   \\
179095  &   A0V   &	 -0.069  &    0.022  &   0.129 & gh   \\
179124  &   B9V   &	  0.280  &    0.095  &   0.146 & d    \\
179483  &   A2V   &	  0.316  &    0.028  &   0.144 & e    \\
179733  &   A0V   &	  0.211  &    0.038  &   0.117 & ac   \\
180138  &   A0V   &	  0.075  &    0.045  &   0.128 & c    \\
180501  &   A0V   &	  0.147  &    0.027  &   0.117 & gh   \\
180681  &   A0V   &	  0.112  &    0.031  &   0.111 & acei \\
183142  &   B8V   &	 -0.462  &    0.060  &   0.093 & ei   \\
184147  &   A0V   &	  0.007  &    0.019  &   0.121 & egi  \\
184787  &   A0V   &	  0.034  &    0.017  &   0.154 & cf   \\
188252  &   B2III &	 -0.461  &    0.047  &   0.155 & ce   \\
188461  &   B3V   &	 -0.461  &    0.109  &   0.095 & efj   \\
189845  &   A0V   &	  0.136  &    0.053  &   0.127 & fj   \\
190025  &   B5V   &	 -0.230  &    0.157  &   0.084 & j    \\
190112  &   A0V   &	  0.067  &    0.027  &   0.113 & f    \\
\hline
\end{tabular} 
\label{tab:cal} 
\end{center}
List of dropped calibrators: HD179395, HD181939, HD182487, 
HD184875, HD189253; ``ID'' refers to the ID of the target star for which the 
calibrator has been used
(see column 3 of Table \ref{tab:pavo}).
\end{small}
\end{table}

\begin{table*}
\begin{small}
\begin{center}
\caption{Interferometric observations and measured parameters.}
\vspace{0.1cm}
\begin{tabular}{l l c c l c c c c c}        
\hline         
HD & KIC  & ID	& Scans/Nights & Baselines 	& $\mu_{R}$	& $\theta_{\rm UD}$ & $\theta_{\rm LD}$ & $\theta_{(V-K)}$ &  $\theta_{\rm IRFM}$   	\\ 
\hline
173701 & 8006161 & a & 9/4     & S2E2,S1W1,S1E2,S1E1  &       $0.59(4)$ & $0.314(4)$  &  $0.332(6)$     	  & $0.333(5)$ & $0.32(1)$	\\
175726 &  -- &     b & 3/2     & S1W2,S1E1     		  &       $0.51(5)$ & $0.331(5)$  &  $0.346(7)$  		  & $0.356(5)$ & $0.35(2)$	\\
177153 & 6106415 & c & 7/3     & S2E2,S1W1,S1E2		  &       $0.51(5)$ & $0.276(6)$  &  $0.289(6)$  		  & $0.285(4)$ & $0.28(1)$		\\
181420 &  -- &     d & 5/2     & S1W2,S1E1     		  &       $0.48(5)$ & $0.32(1)$   &  $0.34(1)$   		  & $0.312(5)$ & $0.31(1)$		\\
182736 & 8751420 & e & 5/4     & S2W2,S2E2,S1W1		  &       $0.59(4)$ & $0.412(3)$  &  $0.436(5)$  		  & $0.429(6)$ & $0.44(2)$	\\
187637 & 6225718 & f & 6/3     & S2E2,S1E2,S1E1       &       $0.49(5)$ & $0.222(5)$  &  $0.231(6)$  		  & $0.222(3)$ & $0.22(1)$		\\
\hline
175955 & 10323222& g & 4/2     & W1W2,S2W2     &       $0.67(2)$ & $0.634(9)$   &  $0.68(1)$   & $0.70(1)$  & $0.66(3)$		\\
177151 & 10716853& h & 4/2     & W1W2,S2W2     &       $0.64(3)$ & $0.541(8)$  &  $0.57(1)$   & $0.57(1)$  & $0.53(2)$		\\
181827 & 8813946 & i & 3/2     & S2W2,S1W1     &       $0.64(3)$ & $0.443(3)$  &  $0.473(5)$  & $0.516(9)$ & $0.49(2)$		\\
189349 & 5737655 & j & 4/3     & S2W2,S1W2     &       $0.58(4)$ & $0.399(4)$  &  $0.420(6)$  & $0.444(9)$ & $0.44(2)$		\\	 
\hline
\end{tabular} 
\label{tab:pavo} 
\end{center}
``ID'' can be used in Table \ref{tab:cal} to identify which stars have been used to 
calibrate this target. All angular diameters are given in units of milli-arcseconds. 
Baselines are sorted from shortest to longest length for a given target.
\end{small}
\end{table*}

Interferometric observations require careful calibration of the observed visibilities. 
Ideally, this is achieved by observing bright, unresolved point sources as closely 
as possible to the target object in time and 
distance. For PAVO observations of targets as small as in our case, this 
means calibrating with
late B to early A stars since at the PAVO magnitude limit these stars are distant enough to 
have significantly smaller diameters (0.1--0.15\,mas) than our target stars. 
Table \ref{tab:cal} lists all calibrators that were used in our analysis. Expected sizes are 
calculated using the 
$V-K$ relation by \citet{kervella04} for dwarf and subgiant stars. 
$V$ band magnitudes have been taken from 
the Tycho catalog and were converted into the Johnson system using the calibration by 
\citet{bessell00}. $K$ magnitudes were adopted from 2MASS \citep{skrutskie06}. 
Interstellar reddening for each calibrator was estimated using the extinction model 
by \citet{drimmel03}.

Although we have 
checked each calibrator in the literature for possible multiplicity, rotation and variability prior to 
observations, our data show that roughly 1/4 of all observed calibrators are more 
resolved than expected, and therefore potentially unsuitable for calibration. 
These calibrators are listed at the bottom of Table \ref{tab:cal}.
Possible reasons for this include previously undetected  
binary systems and rapid rotation causing deviations from spherical symmetry.

\begin{figure*}
\begin{center}
\resizebox{\hsize}{!}{\includegraphics{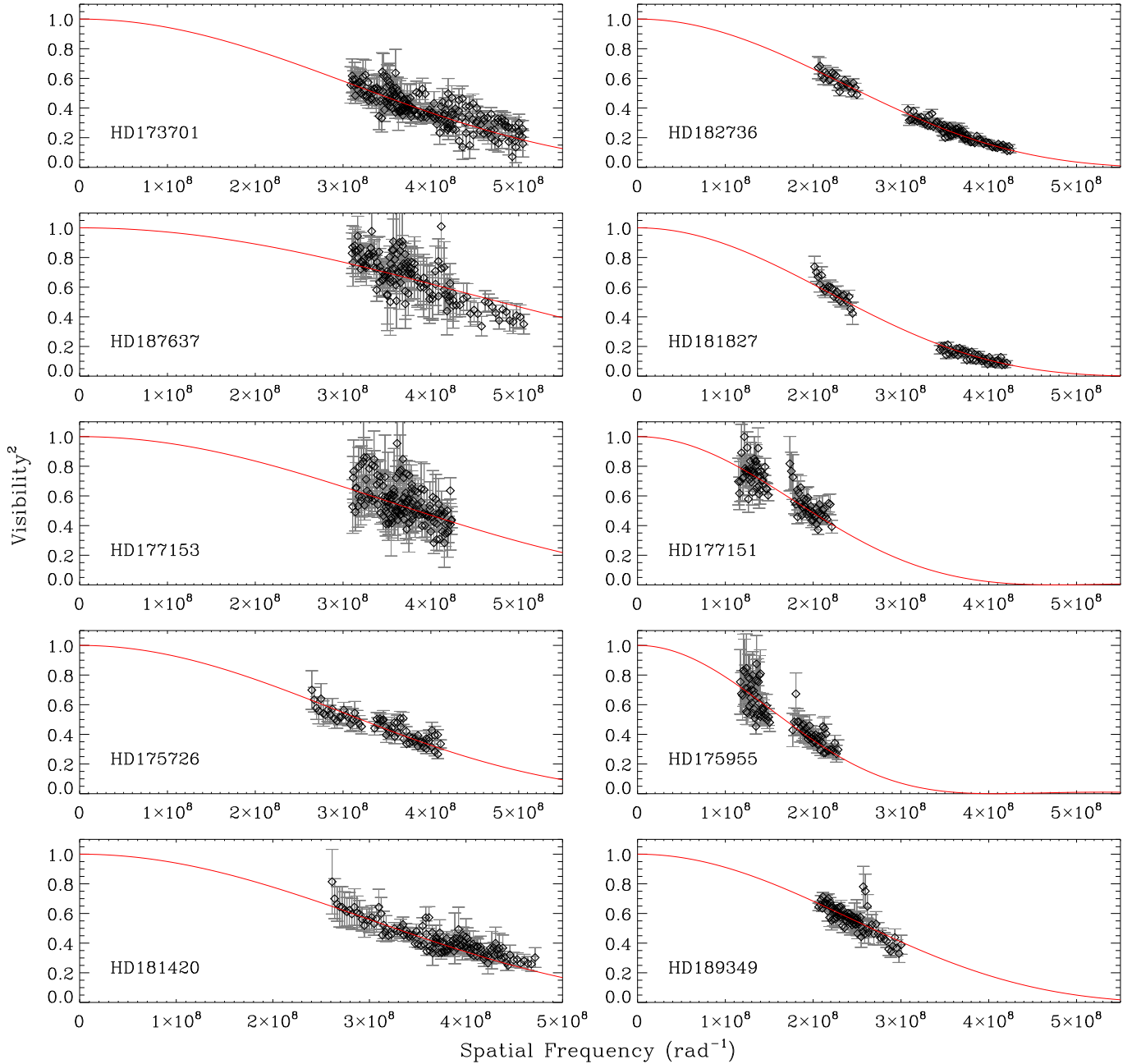}}
\caption{Squared visibility versus spatial frequency for all stars in our sample. Red solid lines 
show the fitted limb-darkened disc model. 
The order of panels is the same as in Figure \ref{fig:ps_chara}. Note that the 
error bars for each star have been scaled so that the reduced $\chi^{2}$ equals unity.}
\label{fig:pavo_chara}
\end{center}
\end{figure*}

Figure \ref{fig:pavo_chara} presents the calibrated squared-visibility measurements as a 
function of spatial frequency for all targets in our sample, with a summary of observations 
given in Table \ref{tab:pavo}. We have collected at least three independent scans for 
each target over at least two different nights, and the visibilities of each target were 
calibrated with at least two different calibrators (see also Table \ref{tab:cal}). 
Note that each scan typically produces a measurement of visibility in 20 independent wavelength 
channels, resulting in a total of $\sim$1000 visibility measurements in our campaign.

For each 
target we fitted the following limb-darkened disc model to the observations \citep{hanbury74b}:

\begin{multline}
V = \left (\frac{1-\mu_{\lambda}}{2}+\frac{\mu_{\lambda}}{3} \right )^{-1} \\ \times \left [(1-\mu_{\lambda})\frac{J_{1}(x)}{x} + \mu_{\lambda} (\pi/2)^{1/2} \frac{J_{3/2}(x)}{x^{3/2}} \right ] \:,
\end{multline}

\noindent
with

\begin{equation}
x = \pi B \theta_{\rm LD} \lambda^{-1} \: .
\end{equation}

\noindent
Here, $V$ is the visibility, $\mu_{\lambda}$ is the linear limb-darkening coefficient, $J_{n}(x)$ is 
the $n$-th order 
Bessel function, $B$ is the projected baseline, $\theta_{\rm LD}$ is the angular diameter after 
correction for limb-darkening, and
$\lambda$ is the wavelength at which the observation was made. Linear 
limb-darkening coefficients in the $R$ band for our 
targets were estimated by interpolating the model grid of \citet{claret11} 
to the spectroscopic estimates of $T_{\rm eff}$, $\log g$ and [Fe/H] (Table \ref{tab:litparas}) 
for a microturbulent velocity of 2\,km\,s$^{-1}$. 
Uncertainties on the limb-darkening coefficients were estimated from the difference in 
the methods presented by \citet{claret11}.
The choice of the limb-darkening model has little effect on the final fitted angular 
diameters. Detailed 3-D hydrodynamical models by \citet{bigot06}, 
\citet{chiavassa10} and \citet{chiavassa12} for dwarfs and giants 
have shown that the differences to simple linear limb darkening models are 
1\% or less in angular diameter for stars with near solar-metallicity. 
For a moderately 
resolved star with $V^{2}\sim 0.5$, a 1\% change in angular diameter would 
arise from a change of less than 1\% in $V^{2}$, which is less 
than our typical measurement uncertainties.

The procedure used to fit the model and estimate the uncertainty in the derived angular diameters 
was described by \citet{derekas11}. In summary, Monte-Carlo simulations 
were performed which took into 
account uncertainties in the adopted wavelength calibration (0.5\%), calibrator sizes (5\%), 
limb-darkening coefficients (see Table \ref{tab:pavo}), 
as well as potential correlations across wavelength channels. 
The resulting fitted angular diameters of each target, corrected for limb-darkening, are given in 
Table \ref{tab:pavo}. We also give the uniform-disc diameters in 
Table \ref{tab:pavo}, which were derived by setting $\mu_{\lambda}=0$ in Equation (1).

A few comments on our derived diameters are necessary. 
Firstly, one calibrator in our sample (HD\,179124), which is the main calibrator for HD\,181420, 
was recently found to be a rapidly 
rotating B star with $v\sin i = 290$\,km/s \citep{lefever10}. This introduces an extra 
uncertainty on the estimated calibrator diameter. 
We have accounted for this by assuming a 20\% uncertainty in the calibrator diameter, 
which roughly corresponds to the maximum change in the average diameter 
expected for rapid rotators \citep{domiciano02}.
Secondly, a few of our target stars (e.g. HD\,187637) are only about 50\% bigger in 
angular size than their calibrators.
This means that the 
uncertainties on the derived diameters will be strongly influenced
by the assumed uncertainties of the 
calibrator diameters, which in our case are 5\%. While such an uncertainty is reasonable compared 
to the scatter in the photometric calibrations \citep[see, e.g.,][]{kervella04}, 
the diameter measurement itself will only be scientifically useful if the uncertainty in the 
measured diameter
is smaller than the precision of indirect techniques. Further data at longer baselines with 
smaller calibrators will be needed to reduce the uncertainties for these targets.

\begin{figure}
\begin{center}
\resizebox{\hsize}{!}{\includegraphics{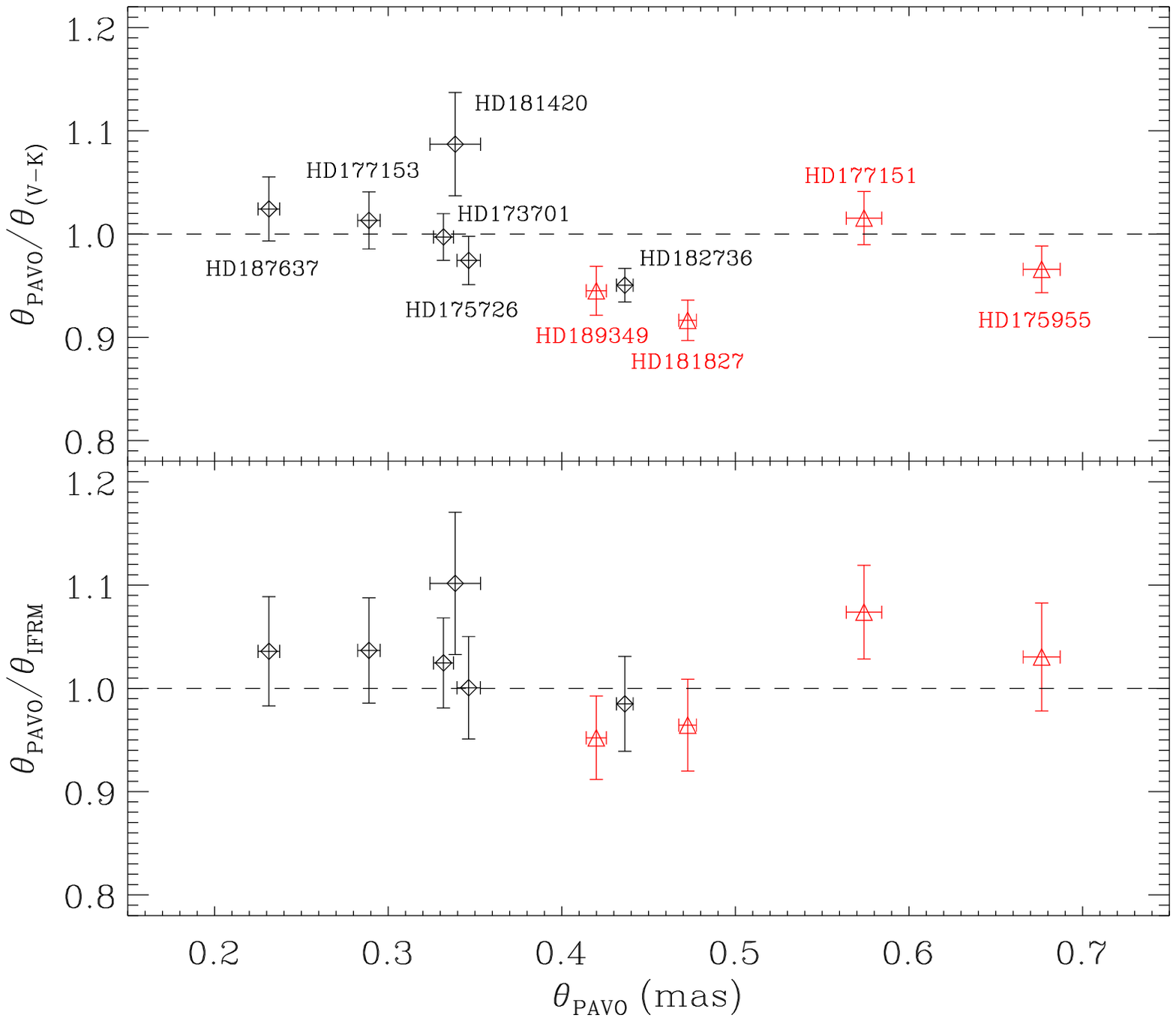}}
\caption{Fractional differences between angular diameters measured with PAVO to diameters 
determined using the $(V-K)$ surface-brightness relation by \citet{kervella04} (upper panel) 
and using the infrared flux method with asteroseismic constraints, as described in 
\citet{silva12} (lower panel). Black diamonds show main-sequence and subgiants stars, 
and red triangles show giant stars. HD numbers of each target are labelled in 
the upper panel.}
\label{fig:size}
\end{center}
\end{figure}

Indirect techniques to estimate angular diameters include surface brightness 
relations \citep[see, e.g.,][]{vanbelle99,kervella04} and the infrared flux method 
\citep[IRFM, see, e.g.,][]{ramirez05,casagrande10}.
Figure \ref{fig:size} compares our measured angular diameters with predictions 
using the $(V-K)$ surface-brightness relation for dwarfs and subgiants 
by \citet{kervella04} and the IRFM method coupled with asteroseismic 
constraints, as described in \citet{silva12}. For the 
$(V-K)$ relation we have adopted a 1\% diameter uncertainty for all stars \citep{kervella04}.
We find good agreement for all stars for both methods, with a 
residual mean of $-2\pm2$\% and $+2\pm2$\% for $(V-K)$ and IRFM, respectively, 
both with a scatter of 5\%. Our results therefore seem to confirm 
that the relation by \citet{kervella04} 
is also valid for red giants, as suggested
by \citet{piau11}, and that combining the IRFM method with asteroseismic 
constraints, as done by \citet{silva12}, 
yields accurate diameters for both evolved and unevolved stars.

These tests of indirect methods are encouraging. We 
emphasize that interferometry remains an important tool to 
validate these methods for a wider 
range of evolutionary states, chemical compositions, and distances. 
The $(V-K)$ relation, for example, 
is based on an empirical relation 
calibrated using nearby stars that does not take into account potential spread 
due to different chemical compositions, and is only valid for de-reddened magnitudes.
An illustration of the importance of using interferometry is 
HD\,181827, which shows a significantly smaller measured diameter than predicted from $(V-K)$. 
This smaller diameter is also in agreement with asteroseismic results, which suggest a 
smaller radius (see Section 4.1).

\subsection{Bolometric Fluxes}
\label{sec:bolflux}
To estimate bolometric fluxes for our target sample, we first extracted synthetic fluxes from 
the MARCS database of stellar model atmospheres \citep{gustafsson08}.
We used models with 
standard chemical composition, with the microturbulence parameter set to 1\,km\,s$^{-1}$ for 
plane-parallel models (unevolved stars) and 2\,km\,s$^{-1}$ for spherical models with a mass of 
1\msun\ for red giants. 
We then multiplied the synthetic 
stellar fluxes by the filter responses 
for the Johnson-Glass-Cousins $UBVRIJHKL$, Tycho $B_{\rm T}V_{\rm T}$ and
2MASS $JHK_{\rm s}$ systems and integrated the 
resulting fluxes to calculate synthetic magnitudes for each MARCS model. 
Filter responses and zeropoints were taken from \citet{bessell12} ($UBVRI$, $B_{\rm T}V_{\rm T}$), 
\citet{cohen03} (2MASS), and \citet{bessel08} ($JHKL$). 
We note that synthetic photometry calculated using MARCS models has previously been 
validated using observed colors in stellar clusters 
\citep{brasseur10,vandenberg10}.
To check the influence of the chosen mass for the spherical models, we have 
repeated the above calculations for typical red giant models with $T_{\rm eff}=5000$\,K 
and $\log g = 2-3$. The fractional differences in the integrated flux for each filter 
for masses ranging from $0.5-5\msun$\ was found to be less than 0.5\% in all bands, and 
are therefore negligible for our analysis.

The amount of photometry in the literature for our sample
is unfortunately small. The targets are generally too faint to have reliable magnitudes in the 
Johnson-Glass-Cousins system, and they are too bright to have a full set of SDSS photometry 
in the KIC. 
To ensure consistency of our bolometric fluxes we only used photometry 
that is available for all stars in our sample, namely Tycho2 $B_{T}V_{T}$ 
and 2MASS $JHK_{\rm s}$ magnitudes. The adopted photometry 
and uncertainties are listed in Table \ref{tab:phot}. 

\begin{figure}
\begin{center}
\resizebox{\hsize}{!}{\includegraphics{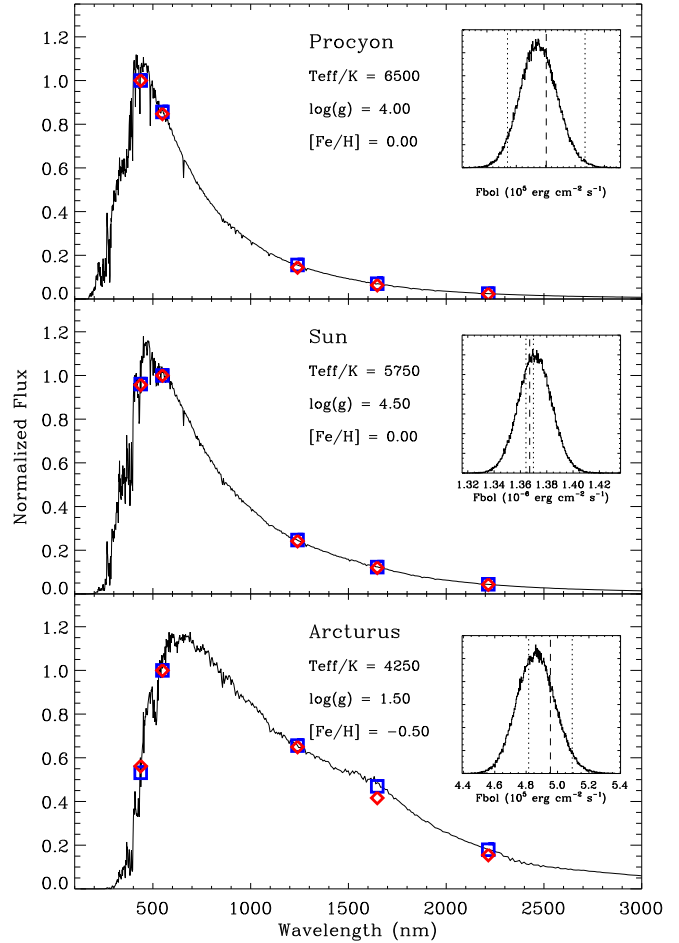}}
\caption{Spectral energy distibutions of Procyon, the Sun and Arcturus to test our method to determine 
bolometric fluxes. 
Black lines are MARCS models with parameters as given in each panel, 
smoothed with a constant spectral resolution $\lambda/\Delta \lambda \sim 200$ 
(corresponding to a width of $\sim$2.5\,nm in the V-band).
Observed and synthetic 
$BVJHK$ photometry is shown as red diamonds and blue squares, respectively.
All fluxes have been normalized to 1 in the band with highest flux for a given star.
The inset shows the result of Monte-Carlo simulations to estimate the uncertainty in the
bolometric flux as described in the text. Dashed and dotted lines show the literature values 
and 1-$\sigma$ uncertainties.}
\label{fig:teststars}
\end{center}
\end{figure}

\begin{figure*}
\begin{center}
\resizebox{\hsize}{!}{\includegraphics{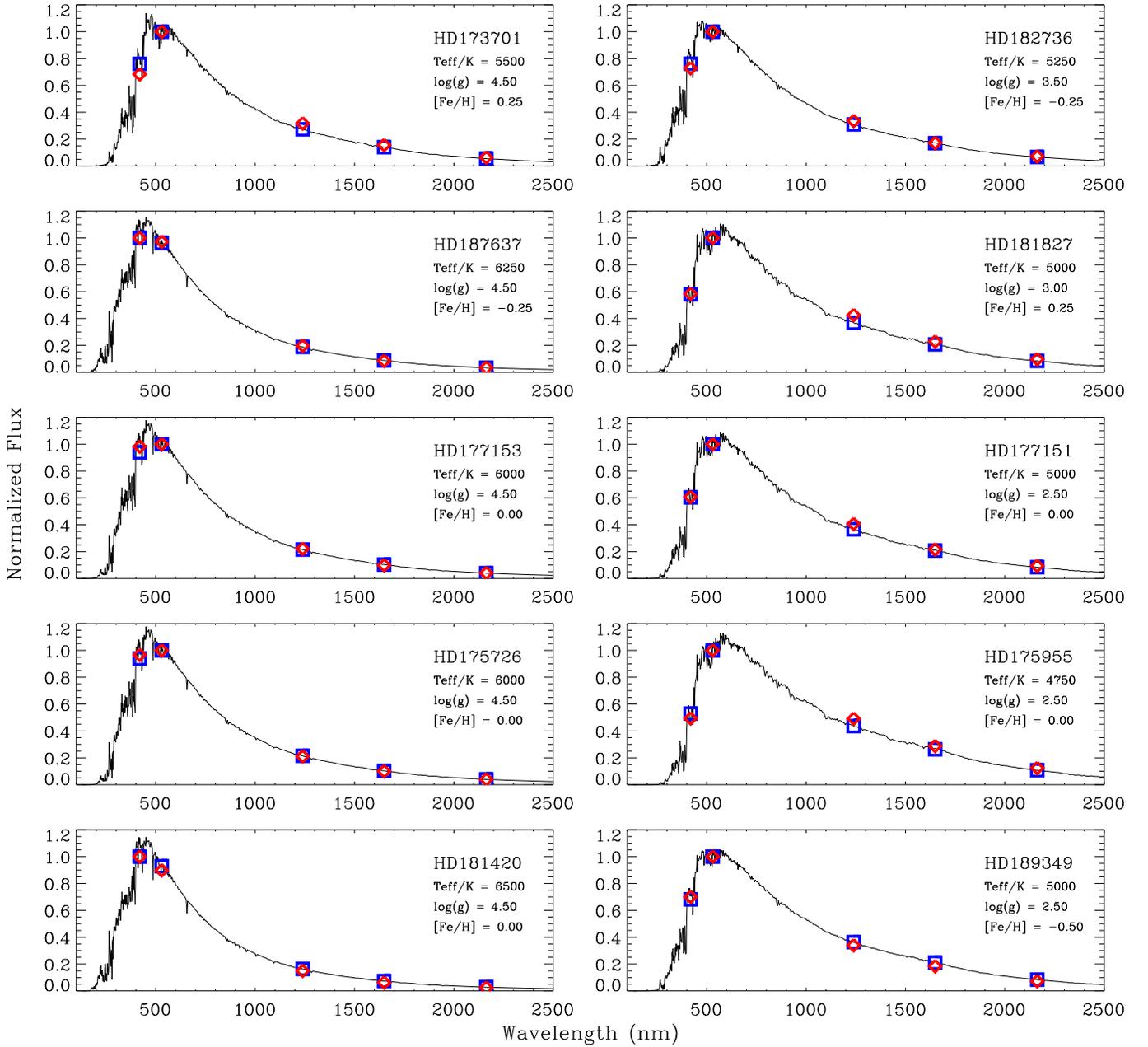}}
\caption{Spectral energy distributions for all stars in our sample. 
Black lines are the MARCS models with parameters as given in each panel, 
smoothed with a constant spectral resolution $\lambda/\Delta \lambda \sim 200$ 
(corresponding to a width of $\sim$2.5\,nm in the V-band). Red diamonds and blue 
squares show the observed and model fluxes in the $B_{T}V_{T}JHK$ bands, respectively.
All fluxes have been normalized to 1 in the band with highest flux for a given star.
The order of panels is the same as in Figure \ref{fig:ps_chara}.}
\label{fig:sed_all}
\end{center}
\end{figure*}

To calculate bolometric fluxes, we largely followed the 
approach described in \citet{alonso95}. For each target star, we first found the six models bracketing 
the spectroscopic determinations $T_{\rm eff}$, $\log g$ and [Fe/H], 
as given in Table \ref{tab:litparas}. We then transformed the 
synthetic $B_{T}V_{T}JHK_{\rm s}$ magnitudes of each model into fluxes, and numerically 
integrated these fluxes using the pivot wavelength for each filter response, calculated
as described by \citet{bessell12} (note that this choice of a reference wavelength is 
independent of the spectral type considered). 
The numerical integration yielded an estimate $f_{\rm int}$, which we then 
compared to the true bolometric flux, $f_{\rm bol}=\sigma T_{\rm eff}^{4}$, where 
$\sigma$ is the Stefan-Boltzmann constant. This yielded 
a correction factor $c=f_{\rm int}/f_{\rm bol}$ for each of the six models, which 
gave the 
percentage of flux included when integrating the photometry over discrete wavelengths. 
The final 
bolometric flux was then calculated by integrating the observed fluxes the same way as the 
model fluxes, and dividing the resulting estimate by the correction factor $c$ found by 
interpolating the six corrections factors to the spectroscopic estimates of 
$T_{\rm eff}$, $\log g$ and [Fe/H]. Note that this interpolation was necessary because the 
step size of the MARCS grid is typically larger than the uncertainties of the 
spectroscopic parameters.
Uncertainties in the derived bolometric fluxes were found by perturbing the input 
photometry and the spectroscopic parameters according to their estimated 
uncertainties (see Tables \ref{tab:litparas} and \ref{tab:phot}), repeating the 
procedure 5000 times, and taking the standard deviation of the resulting distribution.

\begin{table*}
\begin{center}
\caption{Broadband photometry, estimated reddening and bolometric fluxes for all target stars.}
\vspace{0.1cm}
\begin{tabular}{l l c c c c c c c c c}        
\hline         
HD  & KIC &	$B_{\rm T}$ & $V_{\rm T}$ & $J$ & $H$ & $K$ & $E(B-V)$ & \multicolumn{3}{c}{F$_{\rm bol}$ ($10^{-8}$ 
erg s$^{-1}$ cm$^{-2}$)}	\\ 
	&	  &			    &			  &	    &	  &     &          & \scriptsize{MARCS} & \scriptsize{ATLAS+$\theta$} & \scriptsize{ATLAS+$V_{\rm T}$} \\
\hline
173701 &   8006161 &  $8.606(16)$ & $7.610(11)$ & $6.088(21)$ & $5.751(16)$ & $5.670(21)$ & 0.000(5) & $2.89(5)$   & $2.93(4)$ & $2.86(3)$	\\
175726 &   --	   &  $7.401(15)$ & $6.780(10)$ & $5.703(24)$ & $5.418(34)$ & $5.346(20)$ & 0.000(5) & $5.4(1)$   & $5.40(7)$ & $5.41(6)$	\\
177153 &   6106415 &  $7.872(15)$ & $7.275(11)$ & $6.145(20)$ & $5.923(26)$ & $5.829(23)$ & 0.000(5) & $3.39(7)$	 & --        & --		\\
181420 &   --	   &  $7.059(15)$ & $6.604(10)$ & $5.748(21)$ & $5.560(33)$ & $5.513(26)$ & 0.000(5) & $6.0(2)$    & $6.2(1)$  & $5.99(7)$	\\
182736 &   8751420 &  $8.021(16)$ & $7.103(10)$ & $5.515(24)$ & $5.135(27)$ & $5.028(16)$ & 0.000(5) & $4.77(8)$	 & $4.74(5)$ & $4.67(5)$  	\\
187637 &   6225718 &  $8.118(15)$ & $7.580(11)$ & $6.544(21)$ & $6.346(29)$ & $6.283(18)$ & 0.000(5) & $2.55(5)$	 & --        & --		\\
\hline  														       
175955 &   10323222&  $8.513(16)$ & $7.146(10)$ & $4.999(24)$ & $4.442(31)$ & $4.318(17)$ & 0.09(2) & $7.2(3)$	& -- & --	\\	
177151 &   10716853&  $8.281(16)$ & $7.144(10)$ & $5.264(18)$ & $4.820(33)$ & $4.686(20)$ & 0.04(2) & $5.6(3)$	& -- & --	\\	
181827 &   8813946 &  $8.476(15)$ & $7.300(10)$ & $5.453(35)$ & $4.997(15)$ & $4.872(21)$ & 0.04(2) & $4.8(2)$	& -- & --	\\	
189349 &   5737655 &  $8.411(16)$ & $7.411(10)$ & $5.638(24)$ & $5.181(21)$ & $5.124(29)$ & 0.07(2) & $4.6(2)$	& -- & --	\\								
\hline
\end{tabular} 
\label{tab:phot} 
\flushleft Tycho2 and 2MASS photometry are taken from \citet{hog00} and \citet{cutri03}. 
\end{center}
\end{table*}

To test this approach, we have used the same method 
for three bright stars that span a similar range of evolutionary stages as our 
sample and for which bolometric fluxes have been well determined: Procyon, the Sun, and Arcturus. 
Since Tycho and 2MASS photometry is not available for such bright stars, we have 
used $BVJHK$ photometry to mimic the available information for our target sample. Photometry 
has been taken from the General Catalog of Photometric Data \citep[GCDP,][]{mermilliod97} for 
Procyon and Arcturus, and from \citet{colina96} for the Sun.
Figure \ref{fig:teststars} shows the spectral energy distributions (SEDs)  
of all three stars, comparing the MARCS model that best matches the physical parameters of 
each star (black line) to the observed and synthetic fluxes in the $BVJHK$ bands (red and 
blue squares, respectively). Note that the MARCS models have been smoothed to 
a spectral resolution of $\lambda/\Delta \lambda \sim 200$ for better visibility.
The insets show the distributions of the Monte-Carlo simulations described above 
compared to the literature values of the bolometric flux (dashed line) and their 1-$\sigma$ 
uncertainties (dotted lines). Literature bolometric fluxes have been taken from 
\citet{ramirez11} for Arcturus, \citet{aufdenberg05} and \citet{fuhrmann97} for Procyon, and we 
have adopted an effective temperature of $5777\pm3$\,K for the Sun. 
In all three cases, the bolometric flux using our method is recovered within 
1-$\sigma$, with a maximum deviation of $\sim$0.5 $\sigma$ for Arcturus. 

Figure \ref{fig:sed_all} shows the SEDs
of all target stars with the appropriate models for each star, and 
Table \ref{tab:phot} lists our bolometric fluxes based on the procedure described above. 
We note that for the red giants in our sample, interstellar reddening cannot be neglected. 
To estimate reddening using the SED, we adopted the reddening law by \citet{odonnell94} 
\citep[see also][]{cardelli89}
and iterated over $E(B-V)$ to find the observed colors that best fit the colors of the 
six models bracketing the spectroscopic parameters
in Table \ref{tab:litparas}. We then again interpolated to the spectroscopic 
$T_{\rm eff}$, $\log g$ and [Fe/H] values, analogously to the correction factor 
described above. The derived reddening estimates for the 
giants are listed in Table \ref{tab:phot}. 

To further test these results, we have used an independent method to determine bolometric 
fluxes for four stars by combining publicly available flux-calibrated ELODIE spectra \citep{prugniel07}, 
broadband photometry and ATLAS9 models \citep{castelli03,castelli04}.
We started by calculating a grid of ATLAS9 models in the 3-$\sigma$ error box of 
the spectroscopically determined $T_{\rm eff}$, $\log g$ and [Fe/H] 
(see Table \ref{tab:litparas}). Each model spectrum 
was then multiplied by the $B_{\rm T}V_{\rm T}JHK_{\rm s}$ filter passbands and 
integrated over all wavelengths to 
compute a synthetic flux in each band. 
Model fluxes were then
calibrated into fluxes received on Earth using either the 
measured angular diameter or the Tycho $V_{T}$ magnitude.
To find the model that best fits the photometric data we then   
compared the grid of model fluxes with the observed fluxes, calculated using the same 
zeropoints as in the procedure described above.
Finally, the bolometric flux of each star was determined by integrating the ELODIE spectrum between 
390 and 680\,nm together with the synthetic ATLAS9 model (covering
the wavelength ranges $<390$\,nm and $>680$\,nm) that best fits the observed photometry. 

To estimate uncertainties the above procedure was repeated 100 times, drawing 
random values for the observed photometry given in 
Table \ref{tab:phot}, and adding the standard deviation of the resulting distribution 
in quadrature to the 
uncertainty of the total flux of the ELODIE spectra. The final values for the two 
different calibration methods are given 
in Table \ref{tab:phot}. The derived bolometric fluxes agree well with the estimates from 
MARCS models, reassuring us that the model dependency and 
adopted method have little influence compared to the estimated 
uncertainties. We note that we have also compared our bolometric fluxes 
with estimates derived from the infrared flux method, as described in 
\citet{silva12}. Again, we have found good agreement with our estimates within the quoted 
uncertainties.

\section{Fundamental Stellar Properties}

\subsection{Asteroseismic Scaling Relations}
The large frequency separation of oscillation modes with the same spherical 
degree and consecutive radial order is closely related to the mean density of star \citep{ulrich86}:

\begin{equation}
\Delta\nu \propto M^{1/2} R^{-3/2} \: .
\label{equ:dnu}
\end{equation}

\noindent
Additionally, \citet{brown91} argued that the frequency of maximum power (\numax) for solar-like 
stars should scale with the acoustic cut-off frequency, which was used by \citet{KB95} to formulate a 
second scaling relation:

\begin{equation}
\nu_{\rm max} \propto M R^{-2} T_{\rm eff}^{-1/2} \: .
\label{equ:nmax}
\end{equation}

\noindent
Provided the effective temperature of a star is known, 
Equations (\ref{equ:dnu}) and (\ref{equ:nmax}) allow an estimate of the stellar mass and radius. 
This can be done by either combining the two equations 
\citep[the so-called direct method, see][]{kallinger10c} or by comparing the 
observed values of \Dnu\ and \numax\ with values calculated from a grid of evolutionary models 
\citep[the so-called grid-based method, see][]{stello09,basu10,gai11}.

\begin{figure}
\begin{center}
\resizebox{\hsize}{!}{\includegraphics{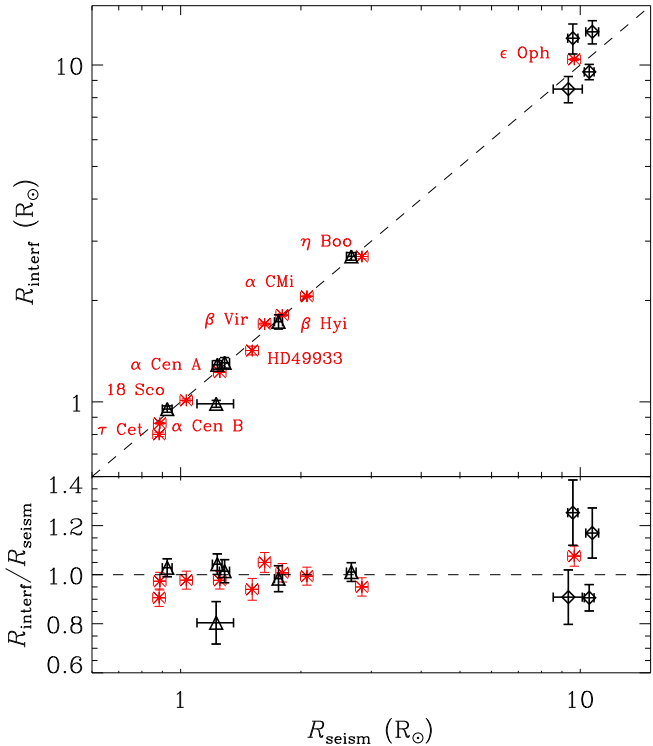}}
\caption{Comparison of stellar radii measured using interferometry and calculated using 
asteroseismic scaling relations. Black diamonds show our \kep\ and CoRoT sample, and red 
asterisks show several bright stars as indicated in the plot for comparison. 
The dashed line marks the 1:1 relation.}
\label{fig:comprad}
\end{center}
\end{figure}

Our interferometric observations, presented in the Section 3.2, allow us to test 
Equations (\ref{equ:dnu}) and (\ref{equ:nmax}).
Using the Hipparcos parallaxes in combination with the angular diameters, we have calculated 
linear radii for our sample of stars, which are listed in Table \ref{tab:finparas}. 
These are compared to asteroseismic radii calculated using Equations (\ref{equ:dnu}) and 
(\ref{equ:nmax}) 
(using $T_{\rm eff}$ values taken from Table \ref{tab:litparas})
in Figure \ref{fig:comprad}. Note the influence of $T_{\rm eff}$ on Equation (\ref{equ:nmax}) 
is small: for solar $T_{\rm eff}$ a variation of 100\,K causes only a 0.9\% change in 
\numax, which is significantly smaller than our typical uncertainties (see Table 
\ref{tab:seism}).

The comparison in Figure \ref{fig:comprad} is very encouraging, showing 
an agreement between the two methods within 3-$\sigma$ in all cases. The overall scatter about the 
residuals is $\sim$13\%, and we do not observe any systematic trend as a function of size (and 
therefore stellar properties). We note that two of the stars in our sample (HD\,173701 and 
HD\,177153) have also been analyzed by \citet{mathur12}, who used both a grid-based approach as well as 
detailed modelling of individual oscillation frequencies to derive stellar radii and masses. 
In both cases, the radii from different models presented in \citet{mathur12} slightly
improve the difference to the interferometrically measured radius, with minimum 
differences of +0.4\,$\sigma$ and +0.8\,$\sigma$ compared to differences of -0.6\,$\sigma$ and 
-1.0\,$\sigma$ from the direct method, respectively.

For comparison, Figure \ref{fig:comprad} also shows examples of bright stars for which 
well constrained asteroseismic and interferometric parameters are available.
We have adopted values for \Dnu\ and \numax\
from \citet{stello09c} and references therein, 
with uncertainties fixed to typical values of 1\% in \Dnu\ and 3\% in \numax. 
Asteroseismic observations 
have been obtained from the MOST space telescope for $\epsilon$\,Oph \citep{barban07,kallinger08a}, 
the CoRoT space-telescope for HD\,49933 \citep{appourchaux08},
and from ground-based Doppler observations for the remaining sample 
\citep{carrier03,kjeldsen03,bedding04,carrier05,carrier05b,kjeldsen05,bedding07,arentoft08,teixeira09,bazot11}.
Angular diameters and 
effective temperatures were taken from \citet{mazumdar09} and \citet{deridder06} for $\epsilon$\,Oph, 
\citet{bazot11} for 18\,Sco, \citet{bigot11} for HD\,49933 
and from \citet{bruntt10} and references therein for the remaining 
sample. Parallaxes were adopted from \citet{leeuwen07}, except for 
$\alpha$ Cen A and B for which we have adopted the value by \citet{soederhjelm99}. 
Figure \ref{fig:comprad} again shows 
agreement within 3-$\sigma$ in all cases. 
Excluding HD\,175726 from our sample due to large uncertainties in the asteroseismic 
observations, the residual scatter between asteroseismic and interferometric radii is 
4\% for dwarfs and 16\% for giants, with mean deviations of $-1\pm1$\,\% and 
$+6\pm4$\,\%, respectively. This is consistent with our observational uncertainties and 
hence empirically confirms that, at least for main-sequence stars, asteroseismic 
radii from scaling relations are accurate to $\lesssim 4\%$. Note that \citet{miglio11} has 
previously found a similar good agreement for a sample of nearby stars, with a 
residual scatter of 6\%.

\begin{figure}
\begin{center}
\resizebox{\hsize}{!}{\includegraphics{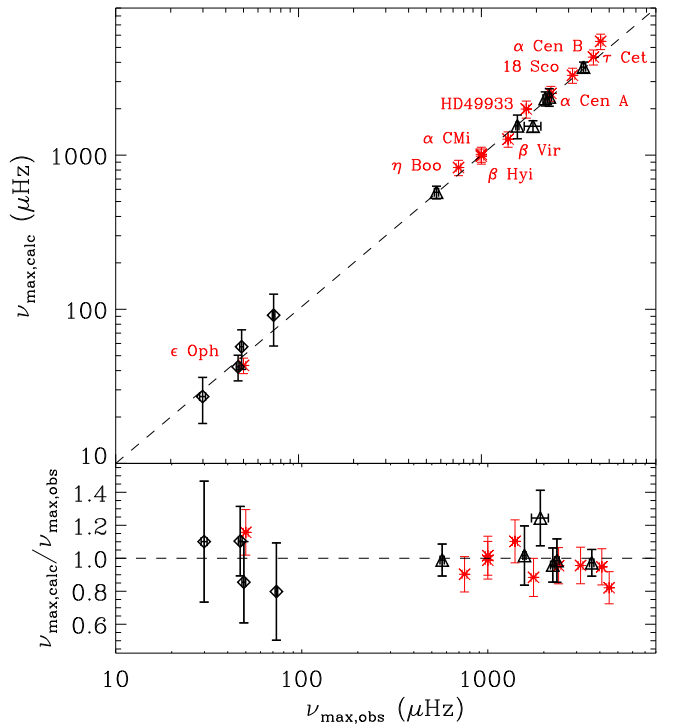}}
\caption{Comparison of \numax\ measured from asteroseismology and calculated using independent 
measurements of R, M and $T_{\rm eff}$. 
Black diamonds show the \kep\ and CoRoT sample, and red 
asterisks show several bright stars as indicated in the plot for comparison.
The dashed line marks the 1:1 relation.}
\label{fig:compnumax}
\end{center}
\end{figure}

It is well known that the scaling relation for \Dnu\ is on more solid ground than the scaling relation 
for \numax, which only recently has been studied in more detail 
observationally \citep[see, e.g.,][]{stello09c,mosser10,white11} 
and theoretically \citep{belkacem11}.
To test Equation (\ref{equ:nmax}), we can combine Equation (\ref{equ:dnu}) with the 
interferometrically measured radii to calculate stellar masses, and combine these 
with $T_{\rm eff}$ to calculate \numax. We 
compare these with the measured values in Figure \ref{fig:compnumax}. We again 
observe good agreement within the error bars, with no systematic deviation as a function 
of evolutionary status. 
Figure \ref{fig:compnumax} also displays a comparison with measured values for a sample of bright stars, again 
showing good agreement with our results for the \kep\ and CoRoT sample. We note that \citet{bedding11b} 
has shown a similar comparison for bright stars, and noted a potential breakdown of the 
\numax\ relation for low-mass stars with $\numax\gtrsim4500\muHz$. Since none of the stars in 
our sample have $\numax>4000\muHz$, we are unable to test this claim in our study.

The large error bars for some stars in Figures \ref{fig:comprad} and \ref{fig:compnumax} may cast 
some doubt about the usefulness of 
interferometry to test scaling relations. Indeed, for the red giants in our sample the 
uncertainty in the interferometric radius is completely dominated by the uncertainty in the parallax. 
For these stars the PAVO data will be most valuable to 
measure the effective temperature by combining the angular diameter with 
an estimate of the bolometric flux, which can then be 
compared to indirect $T_{\rm eff}$ estimates 
from broadband photometry and spectroscopy (see next section). For most unevolved stars in the 
\kep/CoRoT sample, our current uncertainties in the angular diameters are comparable to the 
parallax uncertainties.
The bright star comparison sample, on the other hand, is dominated by the 
uncertainties in the asteroseismic observables, which are much more difficult to constrain from 
the ground or using smaller space telescopes. 
The fact that the asteroseismic uncertainties are almost negligible for the 
\kep/CoRoT sample explains the somewhat counter-intuitive observation that the error bars in 
Figures \ref{fig:comprad} and \ref{fig:compnumax} are similar for some stars of 
the \kep\ sample and for stars which are up to 8 
magnitudes brighter. This comparison underlines the importance of obtaining precise asteroseismic 
constraints on bright stars for which constraints are available from independent 
observational techniques.

\subsection{Spectroscopic and Photometric Temperatures}

The measurement of the angular diameter $\theta_{\rm LD}$ of a star combined with an 
estimate of its bolometric flux $f_{\rm bol}$ allows a direct measurement of the effective 
temperature:

\begin{equation}
T_{\rm eff} = \left (\frac{4 f_{\rm Bol}}{\sigma \theta_{\rm LD}^2}\right)^{1/4} \: ,
\label{equ:teff}
\end{equation}

\noindent
where $\sigma$ is the Stefan-Boltzmann constant. We have used our measured angular diameters 
presented in Section 3.2 together with the bolometric flux estimates presented in Section 3.3 to 
calculate effective temperatures for our sample, which are listed in Table \ref{tab:finparas}.

\begin{figure}
\begin{center}
\resizebox{\hsize}{!}{\includegraphics{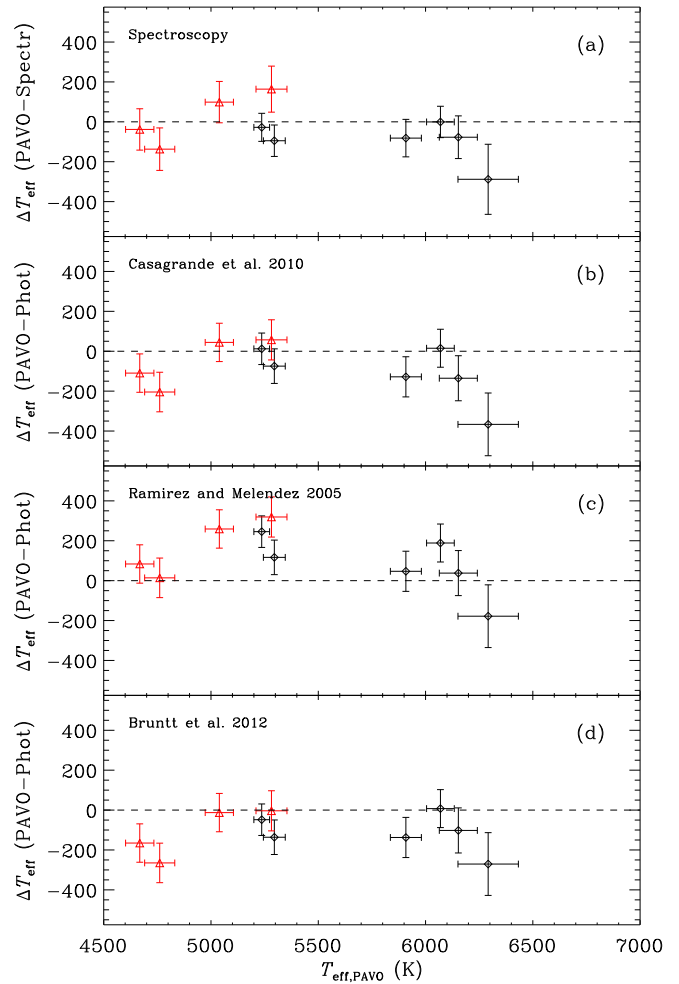}}
\caption{Comparison of effective temperatures derived in this study with spectroscopic 
estimates (panel a) and several photometric calibrations based on $V-K_{\rm s}$ (panels b--d). 
Black diamonds 
are main-sequence stars, while red triangles show red giants. Note that the relations by 
\citet{casagrande10} and \citet{bruntt12} are calibrated for main-sequence stars.}
\label{fig:teff}
\end{center}
\end{figure}

The model dependency of effective temperatures calculated using Equation (\ref{equ:teff}) is small, 
and hence such estimates are important for calibrating indirect photometric estimates such 
as the infrared flux method \citep[see, e.g.,][]{casagrande10}, 
as well as spectroscopic determinations for which usually strong degeneracies between 
$T_{\rm eff}$, $\log g$ and [Fe/H] exist \citep[see, e.g.,][]{torres12}. Figure \ref{fig:teff} compares the measured 
effective temperatures in our 
sample to estimates from high-resolution spectroscopy 
(mostly using the VWA package by \citet{bruntt10}, see Table \ref{tab:litparas}) 
as well as photometric calibrations taken from \citet{casagrande10}, \citet{ramirez05} 
and \citet{bruntt12}. We have chosen $V-K_{\rm s}$ to 
calculate photometric temperatures since this index usually gives the lowest residuals as a 
temperature indicator for cool stars \citep[see, e.g.,][]{casagrande10}. 
The comparison in Figure \ref{fig:teff}a shows good agreement of our 
temperatures with spectroscopy, with a residual mean of $-48\pm39$\,K with a scatter of 
124\,K for all stars, and $-22\pm33$\,K with a scatter of 97\,K when 
excluding the F-star HD\,181420 for which the angular diameter is not well 
determined. We note that this agreement is only slightly worse (with an increased scatter by 
about 10\,K) if we use the $T_{\rm eff}$ values by \citet{bruntt12} and \citet{thygesen12} 
for which no asteroseismic constraints on $\log g$ were used. 
\citet{bruntt10} noted that a slight bias for spectroscopic temperatures to be hotter than 
interferometric estimates by $\sim40\,$K for a sample of nearby stars, which is somewhat 
confirmed by our results, although the scatter is significantly larger. 
Our result confirms that a combination of spectroscopy and asteroseismology 
can be applied for the accurate 
characterization of temperatures, radii and masses of much fainter stars, e.g. 
exoplanet host stars observed by the \kep\ mission.

The photometric estimates shown in Figure \ref{fig:teff}b, \ref{fig:teff}c and 
\ref{fig:teff}d show slight systematic deviations. 
The calibration by \citet{casagrande10} shows the best agreement, with only the 
coolest red giants being slightly hotter than implied by our results. The calibration by 
\citet{ramirez05} is the 
only one that directly provides color-temperature relations 
calibrated for giants. As already noted by \citet{casagrande10}, the temperatures by 
\citet{ramirez05} seem to be systematically cooler than expected, and this is confirmed by 
our results. Finally, the calibration given by \citet{bruntt12} overestimates temperatures at the 
cool end, which is again not surprising since their calibration was based on main-sequence stars 
only, and did not include corrections for lower surface gravities and different metallicities. 
Overall, we conclude that photometric estimates reproduce the measured temperatures 
from interferometry well within the uncertainties, except for the giants where reddening 
is significant. We note that HD\,173701 is the only star with sufficient 
Sloan photometry to be included in the calibration by \citet{pinsonneault11}. The 
SDSS temperature, corrected for metallicity as described in \citet{pinsonneault11}, 
is $5364\pm100$\,K, in good agreement to the determined values 
here. Finally, we note that the effective temperatures presented in this section do not influence 
the comparisons of the asteroseismic masses and radii calculated in the previous 
section (which were calculated using spectroscopic $T_{\rm eff}$), since the dependence of 
Equation (\ref{equ:nmax}) on $T_{\rm eff}$ is only small.

\begin{table*}
\begin{small}
\begin{center}
\caption{Fundamental properties of all Kepler and CoRoT stars in this study.}
\vspace{0.1cm}
\begin{tabular}{l l | c c | c c | c c c}        
\hline         
HD  & KIC & $T_{\rm eff}$ (K) & [Fe/H] & $R/\rsun$ & $M/\msun$ & $R/$\rsun & $M$/\msun &	$T_{\rm eff}$ (K)	\\ 
    &  &\multicolumn{2}{|c|}{Spectroscopy}   & \multicolumn{2}{|c|}{\numax+\Dnu+$T_{\rm eff,sp}$} & $\pi+\Theta_{\rm LD}$ & $\pi+\Theta_{\rm LD}+\Dnu$ & $\Theta_{\rm LD}+f_{\rm bol}$ \\
\hline
173701 &   8006161 & 5390(60) & $+0.34(6)$  & 0.926(26) & 0.969(82)  &  0.952(21) & 1.054(69)    &     5295(51)	       \\
175726 &   --	   & 6070(45) & $-0.07(3)$  & --      & --           &  0.987(23) &	  --       &     6069(65)	       \\
177153 &   6106415 & 5990(60) & $-0.09(6)$  & 1.235(35) & 1.123(94)  &  1.289(37) & 1.27(11)	   &	 5908(72)      \\
181420 &   --	   & 6580(105)& $+0.00(6)$  & 1.758(41) & 1.68(12)   &  1.730(84) & 1.60(23)	   &	 6292(141)	  \\
182736 &   8751420 & 5264(60) & $-0.15(6)$  & 2.674(74) & 1.26(10)   &  2.703(71) & 1.30(10)	   &	 5236(37)	 \\
187637 &   6225718 & 6230(60) & $-0.17(6)$  & 1.288(38) & 1.31(11)   &  1.306(47) &  1.37(15)	   &	  6153(89)	   \\
\hline  	   	      		   	    
175955 &   10323222& 4706(80) & $+0.06(15)$ & 10.53(30)  & 1.51(13)	&  9.54(50)	    & 1.13(18)	&      4668(66)       \\
177151 &   10716853& 4898(80) & $-0.10(15)$ & 10.72(40)  & 1.67(19)	&  12.5(1.0)  & 2.68(64)    &	  4761(70)	      \\
181827 &   8813946 & 4940(80) & $+0.14(15)$ & 9.58(28)   & 2.01(18) 	&  12.0(1.2)  & 4.0(1.2)  &      5039(66)       \\
189349 &   5737655 & 5118(90) & $-0.56(16)$  & 9.34(78)  & 0.79(21) 	&  8.48(76)     & 0.60(17)	&      5282(72)       \\						
\hline
\end{tabular} 
\label{tab:finparas} 
\end{center}
\flushleft  Vertical lines divide estimates based on spectroscopy (columns 3 and 4), 
asteroseismic scaling relations 
only (columns 5 and 6) from estimates using the measured angular diameter 
(columns 7, 8 and 9). No estimates based on asteroseismic constraints 
are reported for HD\,176726 since our results suggest a measurement error for this star (see text).
\end{small}
\end{table*}

\subsection{Stellar Models}

Detailed modelling will be deferred to a future paper, 
but we present some first basic comparisons for the most interesting cases here. 
We use the publicly available BaSTI stellar evolutionary tracks \citep{basti} with 
solar-scaled distribution of heavy elements \citep{grevesse93} and a standard mass loss 
parameter of $\eta=0.4$ \citep[see, e.g.][]{fusipecci76}. 
The models do not include effects of diffusion or gravitational settling, and are 
calibrated to match the observed properties of 
the Sun with a mixing length parameter $\alpha_{\rm MLT}=1.913$ and an
initial chemical composition of $(Y,Z)=(0.2734,0.0198)$. 
No convective-core overshooting was included in the models 
presented here. Note that in the following we 
compare models to radii, masses and temperatures 
derived using our direct measurement of the angular diameter (see columns 7, 8 and 9 
in Table \ref{tab:finparas}) and the spectroscopic metallicities. 

\begin{figure}
\begin{center}
\resizebox{\hsize}{!}{\includegraphics{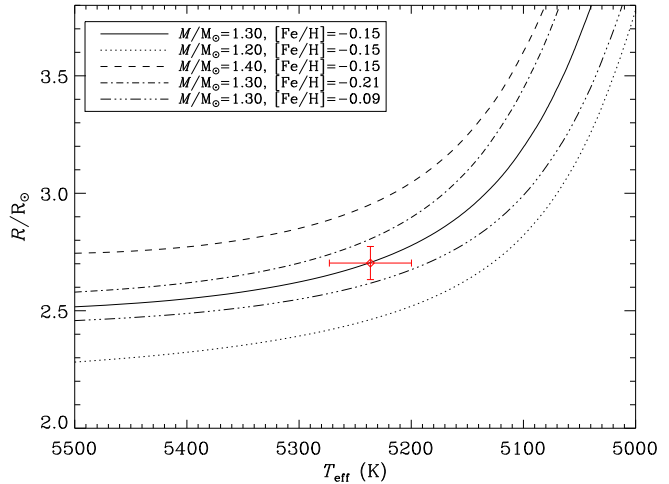}}
\caption{Radius versus effective temperature with the position of the subgiant 
HD\,182736 shown as a red diamond. The solid line shows the BaSTI evolutionary model matching 
the metallicity from high-resolution spectroscopy and the mass determined in this study. 
Dashed-dotted and dashed-triple-dotted lines show the effect of varying the metallicity by 
1\,$\sigma$, while dotted and dashed lines show the same effect for varying the mass by 
1\,$\sigma$. The determined mass and metallicity for HD\,182736 are 
$M=1.3\pm0.1\msun$ and $\rm{[Fe/H]} = -0.15\pm0.06$.}
\label{fig:hd182736}
\end{center}
\end{figure}

\subsubsection{HD\,182736}

The star with the best constrained fundamental properties in our sample is the subgiant 
HD\,182736, with relative uncertainties in temperature, radius and mass of 0.7\%, 2.6\% and 
7.7\%, respectively. The fact that the best observational result is achieved for the only 
subgiant in our sample is not surprising: while the more distant red giants generally have well 
constrained diameters due to their larger size, they suffer from a large uncertainty in the 
parallaxes and effective temperatures due to their larger distance and significant 
reddening. On the other hand, main-sequence stars are generally 
too small to achieve a good precision on their measured diameters. 
Subgiants land in the ''sweet spot'' between these regimes, with 
angular sizes big enough for a precise measurement with PAVO and distances close enough to 
have a well constrained Hipparcos parallax and negligible reddening.

Figure \ref{fig:hd182736} shows a diagram of stellar radius versus effective temperature with 
the position of HD\,182736 according to the properties listed in Table \ref{tab:finparas} 
marked as a red 
diamond. The black solid line shows the evolutionary track matching the determined mass and 
metallicity, calculated by quadratically interpolating the original BaSTI tracks. 
Dashed-dotted and dashed-triple-dotted lines show the effect of varying the metallicity by 
1\,$\sigma$, while dotted and dashed lines show the same effect for varying the mass by 
1\,$\sigma$. The agreement between the models and our observations is excellent, with a 
match within 1\,$\sigma$ for both radius and temperature. We emphasize that no fitting is involved 
in this comparison - the mass, radius, temperature and metallicity are 
determined independently from the evolutionary tracks. A more in-depth asteroseismic 
study using individual frequencies, 
in particular with respect to probing the core rotation rate using mixed modes 
\citep{deheuvels12}, combined with the results presented in this paper should yield 
powerful constraints for studying the structure and 
evolution of this evolved subgiant.

\begin{figure}
\begin{center}
\resizebox{\hsize}{!}{\includegraphics{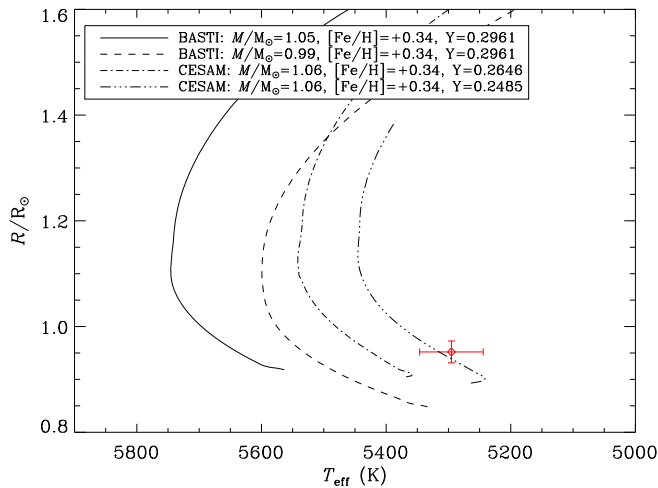}}
\caption{
Radius versus effective temperature with the position of the metal-rich main-sequence star 
HD\,173701 shown as a red diamond. Lines compare BaSTI and CESAM evolutionary tracks with different 
masses and initial Helium fractions (see text). Note that each track starts at the 
zero-age main sequence (ZAMS).
The determined mass and metallicity for HD\,173701 are 
$M=1.05\pm0.07\msun$ and $\rm{[Fe/H]} = +0.34\pm0.06$.}
\label{fig:hd173701}
\end{center}
\end{figure}

\subsubsection{HD\,173701}

Figure \ref{fig:hd173701} shows the radius-$T_{\rm eff}$ diagram for HD\,173701, a metal-rich 
main-sequence star with relatively well constrained properties. In this case, the agreement 
between BaSTI models and observations is poor. 
The difference can be reconciled with a $3-\sigma$ difference 
in mass and metallicity, i.e. the star is more metal-rich and less massive than implied from 
our observations. Indeed, the asteroseismic (but not model-independent) 
analyses 
by \citet{mathur12} and \citet{silva12} imply a mass of $1.00\pm0.01$\,\msun\, and 
$0.96\pm0.04$\,\msun\, for HD\,173701, respectively, which would 
significantly improve the agreement. We also note that the $\sim$100\,K difference to the 
spectroscopic $T_{\rm eff}$ implies that the adopted metallicity may not be consistent with 
the interferometric $T_{\rm eff}$. However, as shown in Figure \ref{fig:hd173701}, 
even at the spectroscopic temperature of 5390\,K the position of HD\,173701 would still 
be slightly too cool for the mass determined from the interferometric radius 
and asteroseismic density. Additionally, adopting a lower $T_{\rm eff}$ in the spectroscopic 
analysis would result in a lower metallicity, and therefore enhance the disagreement 
between models and observations.

A more interesting possibility is that the
physical assumptions in the evolutionary 
models need to be adjusted to reproduce the properties of this star. 
To test this, we have computed additional tracks using the 1D stellar evolution code CESAM 
\citep{morel08}. We use 
opacities from \citet{ferguson05} for the metal repartition by \citet{asplund09}, and 
NACRE nuclear reaction rates are adapted from \citet{angulo99}. The models include 
diffusion and gravitational settling, and
convection is described using the 
mixing length theory by \citet{vitense58} with a solar calibrated value of 1.88.
We have computed two models with the spectroscopically determined 
metallicity of [Fe/H] = 0.34 and a mass of 1.06\msun, once with solar-calibrated initial 
helium mass fraction $Y=0.2646$, and once with $Y=0.2485$, corresponding to the 
lower limit set by cosmological constraints. Figure \ref{fig:hd173701} shows that  
changes in the initial chemical composition brings better agreement to our 
observations. Similar changes can be invoked by reducing the mixing-length parameter 
\citep[see, e.g.,][]{basu10,basu12}. 
\citet{wright04} list 
HD\,173701 with a rotation period of 38 days and Ca H \& K activity of 
$\log(R'_{\rm HK})=-4.87$. Both the slower rotation period and solar-like activity do 
not seem to be compatible with a decreased convection efficiency (smaller mixing length 
parameter), which would be needed to bring 
the models in better agreement to our observations. 
Additionally, a sub-solar helium mass fraction for HD173701 does not seem to be 
compatible 
with the roughly linear helium-to-metal enrichment for metal-rich stars \citep{casagrande07}, 
although the scatter in this relation is large and studies of the Hyades have confirmed that stars can 
be depleted in helium and at the same time have a super-solar metallicity 
\citep{lebreton01,pinsonneault03}.

An alternative explanation could be related to inadequate modeling of 
stellar atmospheres for metal-rich stars. Systematics in these models would affect 
the bolometric flux and hence the determined effective temperature. 
Furthermore, systematic errors in the limb-darkening models for metal-rich 
stars would change the derived angular diameter, which influences the 
determined radius, mass and effective temperature.
Detailed 3-D models by \citet{bigot06} for 
the metal-rich K-dwarf $\alpha$\,Cen\,B showed less significant 
limb-darkening and hence slightly smaller diameters compared to simple 1-D models, 
while \citet{chiavassa10} found differences up to 3\% for models of metal-poor giants. 
Such differences are expected to be enhanced in visibile wavelengths 
(such as the observations presented here) compared to infrared observations 
\citep{allende02,aufdenberg05}. Furthermore, comparisons of 1-D to 3-D 
models have also yielded higher fluxes for 3-D models, particularly at short wavelengths, 
which could lead to small inceases in the derived effective temperature
\citep[see, e.g.,][]{aufdenberg05,casagrande09}.
A higher effective temperature would bring better agreement to the evolutionary tracks and 
spectroscopic estimates. 
More detailed modeling will be 
needed to confirm if refined estimates of limb-darkening, taking into account the non-solar 
metallicity for HD\,173701, can explain 
the observed differences.

\begin{figure}
\begin{center}
\resizebox{\hsize}{!}{\includegraphics{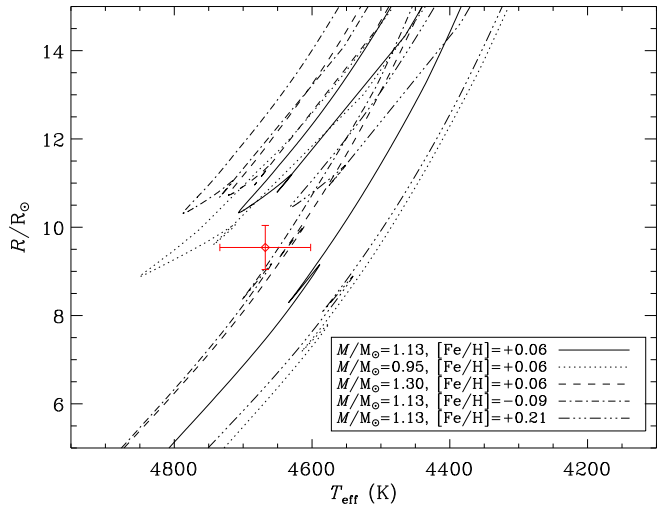}}
\caption{Same as Figure \ref{fig:hd182736} but for the red giant star 
HD\,175955. The determined mass and metallicity for HD\,175955 are 
$M=1.1\pm0.2\msun$ and $\rm{[Fe/H]} = +0.06\pm0.15$.}
\label{fig:hd175955}
\end{center}
\end{figure}

\subsubsection{HD\,175955}

Figure \ref{fig:hd175955} presents a model comparison for HD\,175955, a red giant 
with a well constrained angular diameter and the most precise Hipparcos 
parallax. Gravity mode period spacings measured using 
asteroseismology have been used to classify this star as a H-shell burning, ascending red giant 
branch star \citep{bedding11}. Figure \ref{fig:hd175955} shows that the measured 
temperature of HD\,175955 is slightly hotter than the position of the 
ascending RGB tracks, but overall in good agreement with its determined mass and metallicity. 
Similar to HD\,182736, a combination of the constraints presented here with detailed 
asteroseismic studies \citep[such as the measurement of mixed mode rotational splittings to constrain 
the core rotation rate, see][]{beck12} should allow a detailed 
theoretical study of the internal structure and evolution of this star.

\subsubsection{Additional Notes}

We note that for a few stars the derived stellar properties appear 
unphysical and are likely related to potential observational errors. 
For HD\,175726, for example, the measured linear radius combined with the asteroseismic density 
implies a mass of $0.50\pm0.03$\msun, which seems incompatible with 
its measured radius, temperature and solar-metallicity. 
Using the spectroscopically determined metallicity and the radius and 
temperature from interferometry, a comparison with BaSTI models indicates a mass of 
1.07\msun, which would imply asteroseismic values 
of $\Dnu \sim 142\,\muHz$ and $\numax \sim 3300$\,\muHz. These values are significantly 
different than the results found by \citet{mosser09c}. The difference could be 
explained by an undetected companion causing a significant error in the parallax, 
or due to measurement errors in either the interferometric or the 
asteroseismic analysis. Unfortunately no CoRoT follow-up observations are planned for HD\,175726, 
and hence a resolution of this discrepancy will have to await independent future 
observations.

The red giant HD\,181827, 
on the other hand, has a large mass which is difficult to reconcile with 
evolutionary theory. 
Both the asteroseismic and interferometric constraints are solid, hence 
pointing to a potential problem with the Hipparcos parallax. Indeed, 
HD\,181827 has the largest fractional parallax uncertainty in our sample (10\%), 
leading to a large uncertainty on the radius and hence mass.
We note that HD\,181827 has been asteroseismically identified as a 
secondary clump star \citep{girardi99,bedding11}, corresponding to a 
massive ($\gtrsim 2 \msun$) He-core burning red giant.
Our result of a significantly higher mass for HD\,181827 compared to typical 
red clump giants is hence qualitatively in 
agreement with its asteroseismically determined evolutionary state.

\section{Conclusions}
We have presented interferometrically measured angular diameters of 10 stars for which 
asteroseismic constraints are available from either the \kep\ or CoRoT space telescopes. 
Combining these constraints with parallaxes, spectroscopy and bolometric fluxes, 
we present a full 
set of near model-independent fundamental properties for stars spanning in evolution 
from the main-sequence to the red clump. Our main conclusions from the 
derived properties are as follows:

\begin{enumerate}
\item Our measured angular diameters show good agreement with the surface-brightness relation 
by \citet{kervella04} and the IRFM coupled with asteroseismic constraints by 
\citet{silva12}, with an overall residual scatter of 5\%. Our results seem to confirm 
that the relation by \citet{kervella04} and the method by \citet{silva12} are also reasonably 
accurate for red giants.

\item A comparison of interferometric to asteroseismic radii calculated from scaling relations 
shows excellent agreement within the uncertainties. 
While the uncertainties for giants are large 
due to the uncertainties in the parallaxes, our results empirically prove 
that asteroseismic radii for unevolved stars using simple scaling relations 
are accurate to at least $4\%$. A test of the \numax\ 
scaling relation also shows no systematic deviations as a function of evolutionary state 
within the observational uncertainties.

\item A comparison of measured effective temperatures with estimates 
from modeling high-resolution spectra \citep[mostly using the VWA method, see][]{bruntt10} 
and from the photometric infrared flux method \citep[see][]{casagrande10} 
shows good agreement  with mean deviations of $-22\pm32$\,K 
(with a scatter of 97\,K) and 
$-58\pm31$\,K (with a scatter of 93\,K), respectively, for stars 
between $T_{\rm eff} = 4600-6200$\,K. Some 
photometric calibrations show slight systematic deviations for red giants, presumably 
due to the more significant influence of reddening for these more distant stars.

\item A first comparison of our results with evolutionary models 
shows very good agreement for the subgiant HD\,182736, while there appear to be 
some discrepancies for the metal-rich main-sequence star HD\,173701. We speculate 
that these 
differences may be due to inadequate modeling of stellar atmospheres or limb-darkening 
for metal-rich 
stars, but note that more detailed theoretical studies will be needed to confirm this 
result.
\end{enumerate}

While our study has demonstrated the potential of combining different constraints to test stellar 
model physics, it is clear that the overlap between the different techniques is still 
limited. This situation can be 
expected to be significantly improved with future projects such as the ground-based
network SONG \citep[\textit{Stellar Observations Network Group},][]{grundahl06}, which will 
deliver precise multi-site radial-velocity timeseries for 
asteroseismology and exoplanet studies of nearby 
stars. 
On the other hand, the planned European space mission \textit{Gaia} \citep{perryman} 
will provide accurate parallaxes for stars down to $V<15$, while potential upgrades of interferometers 
such as the CHARA Array with adaptive optics will push the sensitivity 
limits of interferometric follow-up to fainter stars, therefore improving the overlap with
\kep\ and future space-based missions such as TESS 
\citep[\textit{Transiting Exoplanet Survey Satellite},][]{ricker09}. 
The possibility of independently constraining 
radii, effective temperatures, masses and metallicities using asteroseismology, astrometry, 
interferometry and spectroscopy for a large ensemble of stars to study stellar physics
as well as to
characterize potentially habitable exoplanets is clearly the next step for continuing the exciting 
revolution induced by CoRoT and \kep\ 
over the coming decades.

\section*{Acknowledgments}
The authors gratefully acknowledge the \kep\ Science Office and everyone involved in the 
\kep\ mission for making this paper possible. Funding for the 
\kep\ Mission is provided by NASA's Science Mission Directorate. 
The CHARA Array is funded by the National Science Foundation through NSF grant AST-0606958, by 
Georgia State University through the College of Arts and Sciences, and by the W.M. Keck Foundation.
The CoRoT space mission, launched on December 27th 2006, has been developed and is operated 
by CNES, with the contribution of Austria, Belgium, Brazil, ESA (RSSD and Science Programme), 
Germany and Spain. DH is thankful to Karsten Brogaard, Pieter Degroote and Beno\^{i}t Mosser 
for interesting discussions and comments on the paper.
DH, TRB and VM acknowledge support from the Access to Major Research Facilities Program, 
administered by the Australian Nuclear Science and Technology Organisation (ANSTO). 
DH is supported by an appointment to the NASA Postdoctoral Program at Ames Research Center, 
administered by Oak Ridge Associated Universities through a contract with NASA. 
I.M.B. is supported by the grant SFRH / BD / 41213  
/2007 from FCT /MCTES, Portugal. 
J M-\.Z acknowledges the Polish Ministry grant number N N203 405139.
S.G.S acknowledges the support from the Funda\c{c}\~ao para a Ci\^encia e Tecnologia 
(grant ref. SFRH/BPD/47611/2008) and the European Research Council (grant ref. ERC-2009-StG-239953).
WJC acknowledges financial support from the UK Science and Technology 
Facilities Council (STFC).
MC acknowledges funding from FCT (Portugal) and POPH/FSE (EC), for funding through a 
Contrato Ciência 2007 and the project PTDC/CTE-AST/098754/2008.
SH acknowledges financial support from the Netherlands Organisation for Scientific Research (NWO).
KU acknowledges financial support by the Spanish National Plan of 
R\&D for 2010, project AYA2010-17803.
Funding for the Stellar Astrophysics Centre is provided by The Danish  
National Research Foundation. The research is supported by the  
ASTERISK project (ASTERoseismic Investigations with SONG and Kepler)  
funded by the European Research Council (Grant agreement no.: 267864).
This publication makes use of data products from the Two Micron All Sky Survey, which is a joint 
project of the University of Massachusetts and the Infrared Processing and Analysis Center/California 
Institute of Technology, funded by the National Aeronautics and Space Administration and the National 
Science Foundation.

\bibliographystyle{apj}
\bibliography{references}

\end{document}